\documentclass[prd,twocolumn,showpacs,floatfix,superscriptaddress]{revtex4}
\usepackage[utf8]{inputenc}
\usepackage{graphicx}
\usepackage{epsfig}
\usepackage{bm}
\usepackage{amssymb}
\usepackage{float}
\usepackage{amsmath}
\usepackage{dcolumn}
\usepackage{cancel}
\usepackage{graphicx}
\usepackage{epsfig}
\usepackage{bm}
\usepackage[greek,english]{babel} 
\usepackage{alphabeta}
\usepackage{amssymb}
\usepackage{float}
\usepackage{amsmath}
\usepackage{dcolumn}
\usepackage{cancel}
\usepackage[colorlinks]{hyperref}
\usepackage[usenames,dvipsnames]{color}
\hypersetup{
     breaklinks=true,
    pdfstartview={FitH},    % fits the width of the page to the window
    colorlinks=true,       % false: boxed links; true: colored links
    linkcolor=blue,          % color of internal links
    citecolor=red,        % color of links to bibliography
    filecolor=magenta,      % color of file links
    urlcolor=blue,           % color of external links
    anchorcolor=green,      % Color for anchor text
    linktocpage=true
}

\newcommand{\bea}{\begin{eqnarray}}
 
\newcommand{\eea}{\end{eqnarray}}
\newcommand{\bma}{\begin{pmatrix}}
\newcommand{\ema}{\end{pmatrix}}
\newcommand{\be}{\begin{equation}}
\newcommand{\ee}{\end{equation}}
\newcommand{\beno}{\begin{equation*}}
\newcommand{\eeno}{\end{equation*}}

\def\doi{http://doi.org}
\def\arxiv{http://arxiv.org/abs}

\begin{document}

\title{Alleviating $H_0$ tension in Horndeski gravity}

\author{Maria Petronikolou}
\email{petronikoloumaria@mail.ntua.gr}
\affiliation{Department of Physics, National Technical University of Athens, 
Zografou
Campus GR 157 73, Athens, Greece}

\author{Spyros Basilakos}
\email{svasil@academyofathens.gr}
\affiliation{Academy of Athens, Research Center for Astronomy $\&$ Applied 
Mathematics, Soranou Efessiou 4, 11-527, Athens, Greece }
\affiliation{National Observatory of Athens, Lofos Nymfon, 11852 Athens, Greece}
\affiliation{School of Sciences, European University Cyprus, Diogenes Street, 
Engomi, 1516 Nicosia, Cyprus}

\author{Emmanuel N. Saridakis}
\email{msaridak@phys.uoa.gr}
\affiliation{National Observatory of Athens, Lofos Nymfon, 11852 Athens, 
Greece}
\affiliation{CAS Key Laboratory for Researches in Galaxies and Cosmology, 
Department of Astronomy, University of Science and Technology of China, Hefei, 
Anhui 230026, P.R. China}
\affiliation{School of Astronomy, School of Physical Sciences, 
University of Science and Technology of China, Hefei 230026, P.R. China}


%%%%%%%%%%%%%%
\begin{abstract}

We show that the $H_0$ tension can be alleviated in the framework of   
Horndeski/generalized galileon gravity. In particular, since the terms 
depending on $G_5$ control the friction in the Friedmann equation, we 
construct specific sub-classes in which it depends only on the field's kinetic 
energy. Since  the latter is small at high redshifts, namely at redshifts which 
affected 
the CMB structure, the deviations from $\Lambda$CDM cosmology are negligible, 
however as time passes it  increases and thus at low redshifts the Hubble 
function acquires increased values in a controlled way.
We consider two Models, one with quadratic and one with quartic dependence on 
the  field's kinetic energy. In both cases we show the alleviation of the 
tension, resulting to  $H_0 \approx 74$ km/s/Mpc for particular parameter 
choices.
Finally,   we examine the behavior of scalar metric 
perturbations, showing that the conditions for absence of ghost and Laplacian 
instabilities are fulfilled throughout the evolution, and  we 
confront the   models with Supernovae type Ia (SNIa)   and Cosmic 
Chronometer  data.

\end{abstract}

\maketitle

\section{Introduction} 

The Standard Model of Cosmology, namely   $\Lambda$-Cold Dark Matter 
($\Lambda$CDM) plus inflation in the framework of general relativity, proves to 
be very efficient in describing the universe evolution, both at the background 
and perturbation levels \cite{Sahni:1999gb,Peebles:2002gy}. However, theoretical 
issues such 
as the cosmological constant problem and the non-renormalizability of general 
relativity, as well as the possibility of a dynamical nature for the late-time 
acceleration, led to the appearance of various extensions and modifications. In 
general these belong to two classes. In the first class one maintains general 
relativity as the underlying gravitational theory but adds extra components, 
such as the dark energy sectors \cite{Copeland:2006wr,Cai:2009zp}. In the 
second class one constructs modified theories of gravity, which possess general 
relativity as a particular limit but which in general   provide the 
necessary extra degree(s) of freedom that can  drive the universe acceleration
\cite{CANTATA:2021ktz,Capozziello:2011et,Cai:2015emx}.

The last years there have appeared an additional motivation in favour of 
extensions/modifications of the concordance cosmology, namely the need to 
incorporate tensions such as the  $H_0$ and $\sigma_8$ ones. The former arises 
from the fact that   the Planck collaboration estimation for the present 
day cosmic expansion rate  is $H_0 = (67.27\pm 0.60)$ km/s/Mpc 
\cite{Aghanim:2018eyx}, which is 
 in tension at about $4.4$σ with the $2019$ SH0ES collaboration (R19) direct 
measurement, i.e. $H_0 = (74.03\pm1.42)$ km/s/Mpc, obtained using the 
Hubble Space  Telescope observations of $70$ long-period Cepheids in the Large 
Magellanic  Cloud \cite{Riess:2019cxk} (note that  
  combination with gravitational lensing and time-delay data increases the 
deviation at $5.3 \sigma$ \cite{Wong:2019kwg}). Additionally, the $\sigma_8$ 
tension  is related to the   parameter  which 
quantifies the  matter clustering within spheres of  $8 h^{-1} 
\text{Mpc}$ radius, and the  possible deviation   between the Cosmic Microwave 
Background (CMB) estimation 
\cite{Aghanim:2018eyx} and the SDSS/BOSS measurement
\cite{Zarrouk:2018vwy, Alam:2016hwk, Ata:2017dya}.
If these tensions are not a result of unknown systematics, which at least 
concerning the  $H_0$ one seems progressively less possible to be the case, 
then one should indeed seek for alleviation in extensions of the  standard lore 
of cosmology.

In principle one has two main directions to alleviate the  $H_0$ tension. On 
one hand he could alter the  universe content and interactions while 
maintaining general relativity as the gravitational theory 
\cite{DiValentino:2021izs,DiValentino:2020zio,DiValentino:2015ola,  
Bernal:2016gxb,Kumar:2016zpg, 
DiValentino:2017iww,  DiValentino:2017oaw,  Binder:2017lkj, DiValentino:2017zyq,
Sola:2017znb,Yang:2018euj, 
DEramo:2018vss,Poulin:2018cxd,Yang:2018qmz,Yang:2018prh,Pan:2019gop, 
Pan:2019jqh, 
Shafieloo:2016bpk,
Berghaus:2019cls,Pandey:2019plg,Adhikari:2019fvb,Lancaster:2017ksf,
Benisty:2019pxb,
 Perez:2020cwa,Pan:2020bur, 
Benevento:2020fev,Banerjee:2020xcn,Elizalde:2020mfs,Alvarez:2020xmk,
DeFelice:2020cpt, Haridasu:2020pms,
Seto:2021xua,Bernal:2021yli, 
Alestas:2021xes,Elizalde:2021kmo,Krishnan:2021dyb,Theodoropoulos:2021hkk}, 
and on the other hand he could seek for a solution in modified gravity. 
Since the second direction maintains the advantages that modified gravities 
bring related to renormalizability and early- and late-time acceleration, it 
might be preferable. Furthermore,  since  the $H_0$ tension implies that the 
universe expands faster than what $\Lambda$CDM cosmology predicts, in order to 
alleviate it one should seek for a modified gravity that 
qualitatively leads to  ``less gravitational power'' at intermediate 
and late times. Hence, during the last years many models of modified gravity 
have been proposed as candidates for the potential  alleviation of the $H_0$ 
tension 
\cite{Hu:2015rva,Khosravi:2017hfi,Belgacem:2017cqo,El-Zant:2018bsc,
Basilakos:2018arq,Adil:2021zxp,Nunes:2018xbm, 
Lin:2018nxe,Yan:2019gbw,DAgostino:2020dhv,
Anagnostopoulos:2020lec,Capozziello:2020nyq,Saridakis:2019qwt, 
Escamilla-Rivera:2019ulu, DiValentino:2019jae,Vagnozzi:2019ezj,
daSilva:2020bdc,Anagnostopoulos:2019miu,Cai:2019bdh,Braglia:2020auw, 
Abadi:2020hbr,Barker:2020gcp,Wang:2020zfv,Ballardini:2020iws,
LinaresCedeno:2020uxx,Odintsov:2020qzd}.

In this work we are interested in alleviating the $H_0$ tension in the 
framework of  Horndeski gravity.  Horndeski gravity \cite{Horndeski:1974wa}, 
which is 
equivalent to generalized Galileon theory 
\cite{DeFelice:2010nf,Deffayet:2011gz,
Renk:2017rzu},  is the most general  four-dimensional scalar-tensor theory with 
one propagating scalar degree of freedom, that has 
second-order field equations and thus is free  from  Ostrogradski instabilities 
\cite{Ost}. Hence, by choosing suitable sub-classes of the theory we can obtain 
a cosmological behavior that is almost identical with that of $\Lambda$CDM at 
early times, but which at intermediate times deviates from it due to the 
weakening of the gravitational interaction, and thus alleviating the tension 
(see also \cite{Peirone:2019aua,Frusciante:2019puu} for a different approach on 
the problem using cubic covariant Galileon formulation).

 The plan of the work is the following: In Section \ref{galmodel} we present 
 Horndeski gravity, providing the background cosmological equations as well as 
the conditions for pathologies absence at the  perturbation level. In Section 
\ref{Results} we construct specific sub-classes  of  Horndeski gravity that can 
alleviate the  $H_0$ tension, we compare them to $\Lambda$CDM behavior and we 
confront them with  Supernovae type Ia (SNIa) and  Cosmic Chronometer (CC) 
data. Finally, in Section 
\ref{Conclusions} we give a summary of the results and we conclude.

\section{Horndeski gravity}
\label{galmodel}

In this section we briefly review Horndeski gravity, or equivalently 
generalized Galileon theory. We first give the corresponding general action  
and applying it in a cosmological framework we extract the background Friedmann 
equations. Additionally, we give the perturbation equations around such 
background, and we provide  the
conditions for the absence of instabilities.

The most general Lagrangian with one scalar degree of freedom coupled to 
curvature terms, with second-order field equations is
\cite{Horndeski:1974wa,DeFelice:2011bh,Kobayashi:2011nu}
\begin{equation}
{\cal L}=\sum_{i=2}^{5}{\cal L}_{i}\,,\label{Lagsum}
\end{equation}
 with
\begin{align}
&{\cal L}_{2} = K(\phi,X),\label{eachlag2}\\
&{\cal L}_{3} = -G_{3}(\phi,X)\Box\phi,\\
&{\cal L}_{4} = G_{4}(\phi,X)\,
R+G_{4,X}\,[(\Box\phi)^{2}-(\nabla_{\mu}\nabla_{\nu}\phi)\,(\nabla^{\mu}
\nabla^{\nu}\phi)]\,,\\
&{\cal L}_{5} = G_{5}(\phi,X)\,
G_{\mu\nu}\,(\nabla^{\mu}\nabla^{\nu}\phi)\,\nonumber\\&\ \ \
\ \ \ \ -\frac{1}{6}\,
G_{5,X}\,[(\Box\phi)^{3}-3(\Box\phi)\,(\nabla_{\mu}\nabla_{\nu}\phi)\,
(\nabla^{\mu}\nabla^{\nu}\phi)
\,\nonumber\\&\ \ \
\ \ \ \ \ \ \ \ \ \ \ \ \ \ \ \ \ 
+2(\nabla^{\mu}\nabla_{\alpha}\phi)\,(\nabla^
{\alpha}\nabla_{\beta}\phi)\,(\nabla^{\beta}\nabla_{\mu}\phi)]\,.\label{
eachlag5}
\end{align}
In the above expressions $R$ is the Ricci scalar and $G_{\mu\nu}$   the 
Einstein tensor, while the functions $K$ and $G_{i}$ ($i=3,4,5$) depend on the 
scalar field $\phi$
and its kinetic energy
$X=-\partial^{\mu}\phi\partial_{\mu}\phi/2$.
Moreover, 
$G_{i,X}$ and $G_{i,\phi}$ ($i=3,4,5$) denote the
partial
derivatives of $G_{i}$ in terms of $X$ and $\phi$,
i.e. $G_{i,X}\equiv\partial G_{i}/\partial X$ and
$G_{i,\phi}\equiv\partial
G_{i}/\partial\phi$.
 Hence, the total action of the theory will be
\begin{equation}
S=\int d^{4}x\sqrt{-g}\left({\cal L}+{\cal L}_m\right)\,,\label{action1}
\end{equation}
 where $g$  is the metric determinant, and
  ${\cal L}_m$   accounts for the matter
content of the universe, which corresponds to a perfect fluid with energy 
density 
$\rho_m$  and pressure $p_m$.

We consider an expanding Universe  
described by a flat
homogeneous and isotropic Friedmann-Robertson-Walker (FRW) geometry with metric
\begin{equation}
ds^2=-dt^2+a^2(t)\,\delta_{ij} dx^i dx^j,
\label{metric}
\end{equation}
  with $a(t)$  the scale factor.
Varying   the action  (\ref{action1}) with respect to the metric, and imposing 
the above FRW form we obtain the two generalized Friedmann equations: 
\begin{eqnarray}
&&
2X\Kappa_{,X}-\Kappa+6 X\dot{\phi} H G_{3,X}-2X G_{3,\phi}-6H^2 G_{4}
\nonumber\\
&& 
+24H^2X (G_{4,X}+X G_{4,XX}) -12H X \dot{\phi} G_{4,\phi 
X}\nonumber\\
&& -6H \dot{\phi} G_{4,\phi}+2H^3X \dot{\phi} (5G_{5,X}+2X G_{5,XX}
)\nonumber\\
&&\  -6H^2X (3G_{5,\phi}+2X G_{5,\phi X})=-\rho_{m}  ,
\label{Fr1gen}
\end{eqnarray}
\begin{eqnarray}
&&\Kappa-2X (G_{3,\phi}+\ddot{\phi} G_{3,X})+2(3H^2+2\dot{H}
)G_4
\nonumber\\
&&  -8\dot{H}X G_{4,X}
-12H^2 X G_{4,X}-4H\dot{X} G_{4,X}\nonumber\\
&&  
-8H X\dot{X} G_{4,XX}+2(\ddot{\phi}+2H\dot{\phi}) G_{4,\phi}+4X G_{4,
\phi\phi}\nonumber\\
&&  +4X (\ddot{\phi}-2H \dot{\phi}) G_{4,\phi X}-4H^2X^2\ddot{\phi} G_{5,XX}
\nonumber\\
&&  
-2X (2H^3\dot{\phi}+2H\dot{H} 
\dot{\phi}+3H^2\ddot{\phi}) G_{5,X}\nonumber\\
&& 
+4H X(\dot{X}-HX)G_{5,\phi X}+4H X\dot{\phi} G_{5,\phi\phi}\nonumber\\
&&  
+2 [ 2(\dot{H}X+H\dot{X})+3H^2X ] G_{5,\phi}
=-p_m ,
\label{Fr2gen}
\end{eqnarray}
where dots mark derivatives with respect to $t$, and where we have defined the 
Hubble 
parameter 
$H\equiv\dot{a}/a$. Additionally, varying   (\ref{action1}) with 
respect to $\phi(t)$ leads to its
  equation of motion, namely
 \begin{equation}
 \label{scalfieldeq}
\frac{1}{a^3}\frac{d}{dt}(a^3J)=P_{\phi}  ,
\end{equation}
where
\begin{eqnarray}&&J\equiv 
\dot{\phi} \Kappa_{,X}+6H X G_{3,X}-2 \dot{\phi} 
G_{3,\phi}-12H X G_{4,\phi 
X}
\nonumber\\
&&\ \ \ \ \ \  
+6H^2\dot{\phi}(G_{4,X}+2X G_{4,XX})\nonumber\\
&&\ \ \ \ \ \ 
+2H^3X (3G_{5,X}+2X G_{5,XX})\nonumber\\
&&\ \ \ \ \ \ 
+6H^2\dot{\phi} (G_{5,\phi}+X G_{5,\phi X}),
\end{eqnarray}
\begin{eqnarray}&&P_{\phi}\equiv 
\Kappa_{,\phi}-2X (G_{3,\phi\phi}+\ddot{\phi} G_{3,\phi 
X})\nonumber\\
&&\ \ \ \ \ \ \    +6 (2H^2+\dot{H}) G_{4,\phi}
+6H (\dot{X}+2H X) G_{4,\phi X}\nonumber\\
&&\ \ \ \ \ \ \     -6H^2X G_{5,\phi\phi}+2H^3X \dot{\phi} G_{5,\phi 
X} .
\end{eqnarray} 
Finally, the system of equations closes by considering the matter conservation 
equation
\begin{eqnarray}
\dot{\rho}_m+3H(\rho_m+p_m)  =  0.
\label{rhoconsrv}
\end{eqnarray}

Having obtained the background equations of motion, one can proceed to the 
investigation of   perturbations 
\cite{DeFelice:2010pv,DeFelice:2011bh,Appleby:2011aa}. In this work we are 
interested in the scalar perturbations, and specifically on the conditions of 
absence of   ghosts and Laplacian instabilities, in order to ensure that our 
solutions are  cosmologically
viable. In particular, in order for Horndeski/generalized Galileon theory to be 
free from Laplacian instabilities
associated with the scalar field propagation speed one should have
\cite{DeFelice:2011bh}
\begin{equation}
c_{S}^{2}\equiv\frac{3(2w_{1}^{2}w_{2}H-w_{2}^{2}w_{4}+4w_{1}w_{2}\dot{w}_{1}
-2w_{1}^{2}
\dot{w}_{2})
}{w_{1}(4w_{1}w_{3}+9w_{2}^{2})}
\geq0.
\label{cscon}
\end{equation}
Similarly, for the absence of perturbative ghosts one should have 
 \cite{DeFelice:2011bh}
  \begin{equation}
Q_{S}\equiv\frac{w_{1}(4w_{1}w_{3}+9w_{2}^{2})}{3w_{2}^{2}}>0 .
\label{Qscon}
\end{equation}
 In the above expressions we have set
\begin{eqnarray}
&&
\!\!\!\!\!\!\!
w_{1}  \equiv  2 (G_{{4}}-2 
XG_{{4,X}})-2X (G_{{5,X}}{\dot{\phi}}H-G_{{5,\phi}})
 ,
\label{w1def}\\
&&
\!\!\!\!\!\!\!
w_{2}  \equiv  -2  G_{{3,X}}X\dot{\phi}+4  G_{{4}}H-16 {X}^{2}G_{{4,{
XX}}}H
\nonumber
\\
 &  & \ \ \  \ \ 
 +4(\dot{\phi}G_{
{4,\phi X}}-4H  G_{{4,X}})X+2  G_{{4,\phi}}\dot{\phi}
\nonumber
\\
 &  &\ \ \  \ \ 
 +8 {X}^{2}HG_{{5,\phi X}}+2H  X (6G_{{5,\phi}}-5 
G_{{5,X}}\dot{\phi}{H})
\nonumber
\\
 &  & \ \ \  \ \ 
 -4G_{{5,{
XX}}}{\dot{\phi}}X^{2}{H}^{2} ,\\
&&\!\!\!\!\!\!\!
w_{3}  
\equiv  3  X(K_{,{X}}+2  XK_{,{  XX}})\nonumber
\\
 &  &
 \   +6X\!\left(3X\dot{\phi}HG_{{3,{
XX}}}\!-\!G_{{3,\phi X}}
X\!-\!G_{{3,\phi}}\!+\!6  H\dot{\phi}G_{{3,X}}\right)
\nonumber \\
 &  &   \   
 +18  H\!\left(4  H{X}^{3}G_{{4,{  XXX}}}\!-\!5  X\dot{\phi}G_{{4,\phi
X}} \!+\!7  HG_{{4,X}}X\right.\nonumber
\\
 &  &
 \left.\ \ \   \  \
 -\!HG_4 \!-\!G_{{4,\phi}}\dot{
\phi}\!+\!16  H{X}^{2}G_{{4,{  
XX}}}\!-\!2 {X}^{2}\dot{\phi}G_{{4,\phi{
XX}}}\right)\nonumber \\
 &  &   \     
 +6{H}^{2}X\!
 \left(2  H\dot{\phi}G_{{5,{
XXX}}}{X}^{2}\!-\!6 {X}^{2}G_{{5,\phi{  XX}}}\!-\!18G_{{5,\phi}}
\right.\nonumber\\
&&\left.\ \ \   \  \
+13XH\dot{\phi}G_{{5,{  
XX}}}\!-\!27G_{{5,\phi
X}}X\!+\!15  H\dot{\phi}G_{{5,X}}\right),\\
&&\!\!\!\!\!\!\!
w_{4}  \equiv  2G_{4}-2XG_{5,\phi}\!-\!2XG_{5,X}\ddot{\phi}~.
\label{w4def}
\end{eqnarray}
We mention here that  a negative sound speed square should be definitely 
avoided, however a sound speed square larger than one
does not necessarily imply 
pathologies and acausal behavior \cite{Babichev:2007dw,Deffayet:2010qz}.

Lastly, we mention here that in Horndeski theories the  gravitational-wave 
speed is in general   different than 1, namely than the light speed. In 
particular, we have \cite{DeFelice:2011bh}
\begin{equation}
c_{T}^{2}\equiv\frac{w_4 }{w_1}
\geq0,
\label{cTcon2}
\end{equation}
and as we can see from (\ref{w1def}),(\ref{w4def}) the $G_5$ terms may have an 
effect according to the cosmological evolution.

\section{Alleviating the $H_0$ tension}
\label{Results}

In the previous section we presented the cosmological equations in the 
framework of Horndeski/generalized Galileon gravity. In this section we desire 
to use particular sub-classes of the theory in order to obtain an 
alleviating of the $H_0$ tension. Our strategy is the following: since the 
simplest model in Horndeski cosmology is $\Lambda$CDM one, arising from  
$G_4=1/(16\pi G)$, $K=-2\Lambda=const$, and $G_3=G_5=0$,  we want to introduce 
deviations which will be negligible at high redshifts, in which CMB structure 
is 
formed, but that will play a role at low redshifts, in which direct Hubble 
measurements take place. In particular, since it is known that the terms 
depending on $G_5$   affect the   friction 
term on the scalar field 
\cite{Saridakis:2010mf,Capozziello:1999uwa,Koutsoumbas:2013boa,Feng:2013pba,
Koutsoumbas:2017fxp,MohseniSadjadi:2013iou,Dalianis:2016wpu,Karydas:2021wmx}, 
we could consider $G_5$ functions depending only on the kinetic energy $X$ in a 
way that their effect is negligible at high redshifts while being gradually 
important in a controlled way at low redshifts.

Having these in mind, in the following we will consider $G_4=1/(16\pi G)$ and 
$G_3=0$, which are also the case in $\Lambda$CDM cosmology, we will impose a 
simple scalar field potential and standard kinetic term, hence  $K= 
-V(\phi) + X$, and we will  consider 
  the $G_5$ term to depend only on $X$, namely $G_5(\phi,X)=G_5(X)$.
In this case, the Friedmann equations (\ref{Fr1gen}),(\ref{Fr2gen}) become 
\begin{eqnarray}
\label{FR1}
&&
H^{2}=\frac{8\pi G}{3}\Big(\rho_{DE}+\rho_{m}\Big),
\\
\label{FR2}
&&
\dot{H}=-4\pi G \Big(\rho_{DE}+p_{DE}+\rho_{m}+p_{m}
\Big).
\end{eqnarray}
In these equations we have defined an  effective dark energy sector with 
energy density
and pressure respectively:
\begin{eqnarray}
\label{rhode}
&&
\!\!\!\!\!\!\!\!\!\!
\rho_{DE}= 2X-K+2H^3X\dot{\phi}(5G_{5,X}+2X G_{5,XX}) ,\\
\label{pde}
&&\!\!\!\!\!\!\!\!\!\! p_{DE}= K-2X G_{5,X}\left(
2H^3\dot{\phi}+2H\dot{H}\dot{\phi}+3H^2 \ddot{\phi}\right)
\nonumber
\\  
 &  &  \ \ \ \ 
 -4H^2 
X^2\ddot{\phi}G_{5,XX},
 \end{eqnarray}
 and thus the dark-energy equation-of-state parameter becomes
 \begin{eqnarray}
\label{wde1}
w_{DE}\equiv\frac{p_{DE}}{\rho_{DE}}. 
 \end{eqnarray}
 
 Note that the scalar-field conservation equation (\ref{scalfieldeq}) becomes 
simply 
\begin{eqnarray}
\dot{\rho}_{DE}+3H(\rho_{DE}+p_{DE})  =  0.
\label{rhoconsrvDE}
\end{eqnarray}

As we mentioned above we want to make our model to coincide with $\Lambda$CDM 
cosmology at high redshifts. Thus, it proves convenient to use the redshift 
$z=-1+a_0/a$ as the independent variable, fixing the current scale factor  
$a_0=1$ (therefore  $\dot{H}= -(1+z)H(z)H'(z)$ where primes denote 
derivatives with respect to $z$). Introducing as usual the     
  matter density parameter  through 
$ \Omega_m\equiv\frac{8\pi G\rho_m}{3H^{2}}  
$, 
we can express the Hubble function in the case of $\Lambda$CDM 
cosmology as 
\begin{eqnarray}
 H_{\Lambda \text{CDM}}(z) \equiv H_0 
\sqrt{\Omega_{m0}(1+z)^3+1-\Omega_{m0}},
\label{HLCDM1}
\end{eqnarray}
with $H_0$ the Hubble parameter at present and $\Omega_{m0}$ the present value 
of the matter density parameter.

Hence, we want to suitably choose $G_5(X)$ forms in order for the $H(z)$ 
obtained from (\ref{FR1}),(\ref{rhode}) to coincide with  $H_{\Lambda 
\text{CDM}}(z)$ of  (\ref{HLCDM1}) at $z= z_{\rm CMB}\approx 1100$,
namely $H(z\rightarrow z_{\rm CMB}) \approx H_{\Lambda\text{CDM}}(z\rightarrow 
z_{\rm CMB})$,
but give 
$H(z\rightarrow 0) > H_{\Lambda\text{CDM}}(z\rightarrow 0)$.
In the following subsections we will consider  
two sub-cases of the $G_5(X)$ term separately. For simplicity, from now on we 
  focus on the dust matter case, i.e. we impose $p_m=0$, while for the 
scalar-potential without loss of generality we choose $K= -V_0\phi 
+ X$.

\subsection{Model I: $G_5(X)=\xi X^{2}$}

The first model we consider is the one with  $G_5(X)=\xi X^{2}$, 
i.e $G_5$ has a quadratic   dependence on the field's kinetic energy. In this 
case (\ref{rhode}) and (\ref{pde})  respectively become
\begin{eqnarray}
\label{rhode2}
&&
\!\!\!\!\!\!\!\!\!\!\!\!\!\!\!\!
\rho_{DE}=  \frac{\dot{\phi}^2}{2}+V_0\phi+7\xi H^3\dot{\phi}^5,\\
\label{pde2}
&&\!\!\!\!\!\!\!\!\!\!
\!\!\!\!\!\!p_{DE}=  \frac{\dot{\phi}^2}{2}-V_0\phi-\xi\dot{\phi}^4 \left( 
2H^3\dot{\phi}+2H\dot{H}\dot{\phi}+5H^2\ddot{\phi}\right) .
 \end{eqnarray}
As described above we chose the model parameter $V_0$ and the initial 
conditions for the scalar field in order to obtain $H(z_{\rm CMB}) 
=H_{\Lambda\text{CDM}}( z_{\rm CMB})$ and $\Omega_{m0}=0.31$ in agreement with 
 \cite{Planck:2018vyg}, and we leave $\xi$ as the parameter that determines 
the late-time deviation from $\Lambda$CDM cosmology. 

\begin{figure}[H]
%\hspace{-0.3cm}
	\includegraphics[width=0.5\textwidth]{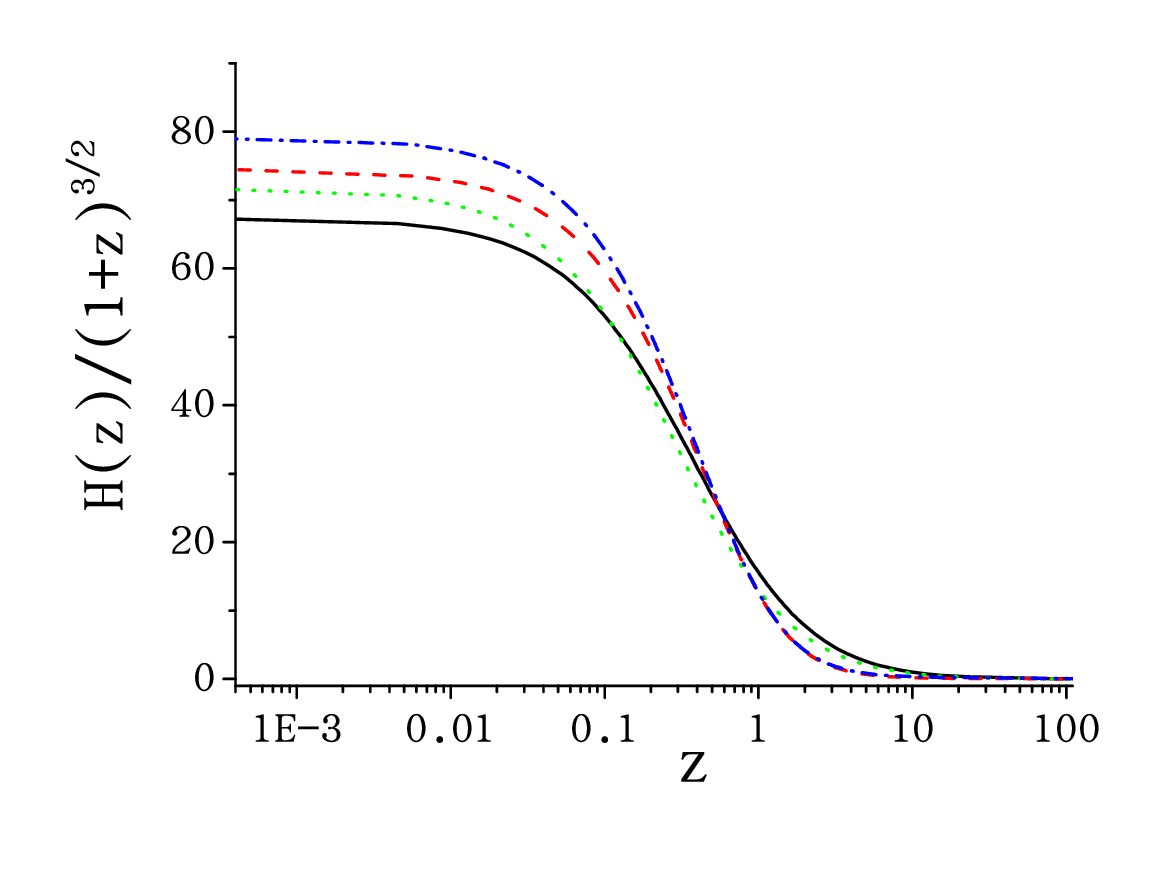}
	\caption{{\it{
	The normalized $H(z)/(1+z)^{3/2}$ in units of km/s/Mpc  as a 
function of the redshift, for $\Lambda$CDM cosmology (black - solid) and for   
Model I  with $V_0=0.08$   and with $G_5(X)=\xi X^{2}$, for $\xi=1.5$ 
(green - dotted),  $\xi=1.3$ 
(red - dashed)
 and  $\xi=1$ (blue - dashed-dotted), in  $H_0$  units. We have imposed 
$\Omega_{m0}\approx0.31$.
}}}
	\label{xiX2}
\end{figure}

In Fig.  \ref{xiX2} we depict the normalized $H(z)/(1+z)^{3/2}$ as a 
function of the redshift, for $\Lambda$CDM cosmology and for our model with 
various choices of 
$\xi$. 
As we can see, indeed our model coincides with $\Lambda$CDM cosmology at high 
and 
intermediate redshifts, while at small redshifts the proposed Model I gives 
higher values. In particular, the present-day value $H_0$ depends on the model 
parameter $\xi$, and it can be around $H_0 \approx 74$ 
km/s/Mpc for $\xi=1.3$ (in   $H_0$ units, i.e. where the $\Lambda$CDM $H_0$ 
is 1).  Specifically, the tension can be alleviated at 3$\sigma$ if 
$1.2<\xi<1.7$.    Hence, we can see that this 
particular sub-class of
Horndeski/generalized Galileon gravity can alleviate the $H_0$ tension due to 
the effect of the kinetic-energy-dependent $G_5$ term. Specifically, at early 
times the field's kinetic term is negligible and hence the  $G_5(X)$ terms   
  do not introduce any deviation from $\Lambda$CDM scenario, however as time 
passes they increase in a controlled and suitable way in order to make the 
Hubble function, and thus $H_0$ too, to increase. Note that, since 
$H_0\approx 10^{-61}$ in Planck  units, the fact that  $V_0=0.08$ and 
$\xi=1.3$ in   $H_0$ units  implies that $V_0\approx 0.5 \times 
10^{-61}$ and $\xi\approx 8 \times 
10^{365}$ in Planck  units (in Planck  units  we obtain   characteristic values 
of $\dot{\phi}$ and  $\phi$   around  $10^{-60}$),
which is the expected scale for the 
quantities  of a scenario that describes the Universe acceleration ($\xi$ 
has dimensions of $[M]^{-9}$ i.e. $\xi^{1/9}\sim 10^{40}$GeV$^{-1}$).

\begin{figure}[H]
% \hspace{-0.3cm}
	\includegraphics[width=0.45\textwidth]{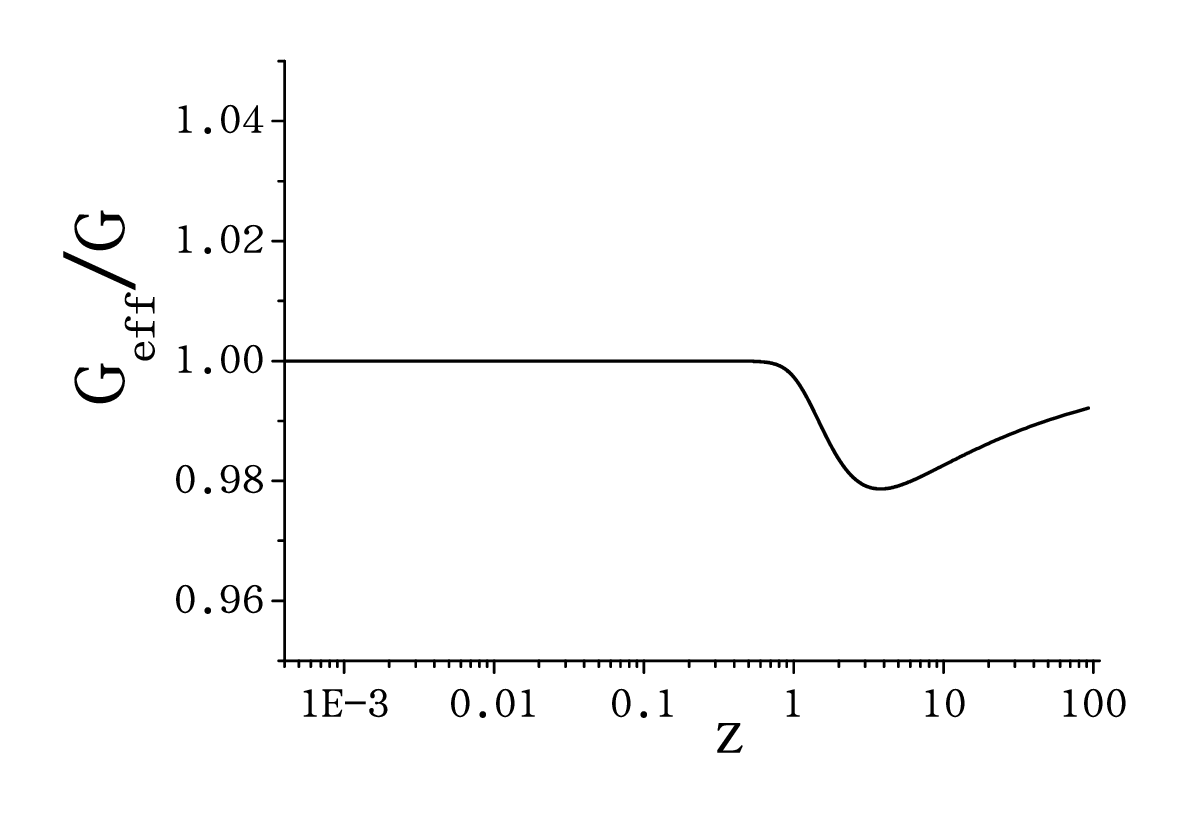}
	\caption{{\it{The normalized effective Newton's constant $ 
\frac{G_{eff}}{G}$ given in 
(\ref{GeffG}) as a function of the 
redshift, for Model I   with $V_0=0.08$  and with $\xi=1.3$ in  $H_0$  
units. 	 	
}}}
	\label{Gefffig}
\end{figure}

\begin{figure}[H]
% \hspace{-0.3cm}
	\includegraphics[width=0.45\textwidth]{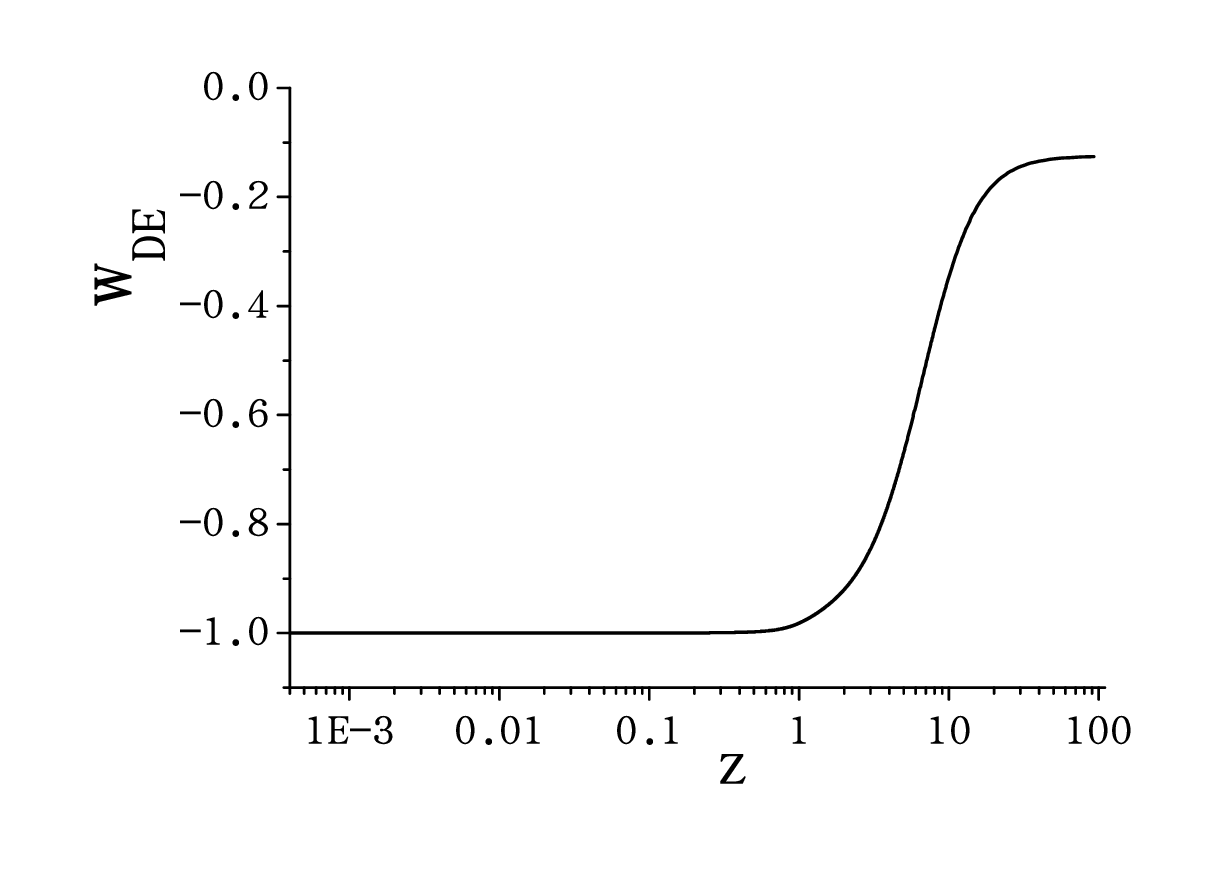}
	\caption{{\it{
The  effective dark-energy equation-of-state parameter $ 
w_{DE}$ given in 
  (\ref{wde1})  as a function of the 
redshift, for Model I   with $V_0=0.08$  and with $\xi=1.3$ in  $H_0$  
units. 	 	 
}}}
	\label{wDEModelI}
\end{figure}

Let us make a comment here on the specific mechanism behind the tension 
alleviation. In general, the alleviation of the $H_0$ 
tension or/and the $\sigma_8$ tension, is a complex issue, and it usually 
arises 
  as a collective result of many effects. If one remains in the class of 
late-time modification (without examining possible early-time solutions, as it 
is the aim of this work) then one efficient mechanism is to have $w_{DE}<-1$ 
at some recent redshift, since a ``phantom'' dark energy implies ``faster'' 
expansion. Nevertheless, this requirement is efficient but not necessary, since 
 the decrease of the effective
Newton's constant at intermediate redshifts is also an efficient mechanism 
\cite{Yan:2019gbw,Heisenberg:2022lob} (see also the   discussion in the recent 
review \cite{Abdalla:2022yfr}), since   ``weaker'' gravity implies ``faster'' 
expansion. In the models proposed in the present work, although many terms are 
involved,  the alleviation of the 
tension arises from such a decreased effective Newton's constant, brought 
about in turn by the friction term. In particular,   in Horndeski theories 
we have \cite{Bellini:2014fua,Peirone:2017ywi}
\begin{equation}
\frac{G_{eff}}{G}=\frac{1}{2}\left(G_4-2X G_{4,X}+X 
G_{5,\phi}-\dot{\phi}HXG_{5,X}
 \right)^{-1},
\label{GeffG}
\end{equation}
and for the present Model I this exhibits a decrease   as 
can be seen in Fig. \ref{Gefffig}. Hence, although $w_{DE}$ in this scenario 
does not evolve into the phantom regime, as we show in Fig. \ref{wDEModelI} 
(note that contrary to tracking dark-energy models 
\cite{Steinhardt:1999nw}, $w_{DE}$ remains strictly negative for $z\gg 1$, i.e.  
 for $100\leq 
z<1000$  the $w_{DE}$  is around $-0.13$), 
the aforementioned decrease in the effective Newton's constant is adequate to 
alleviate the tension.


\begin{figure}[H]
	\centering
	\includegraphics[width=0.44\textwidth]{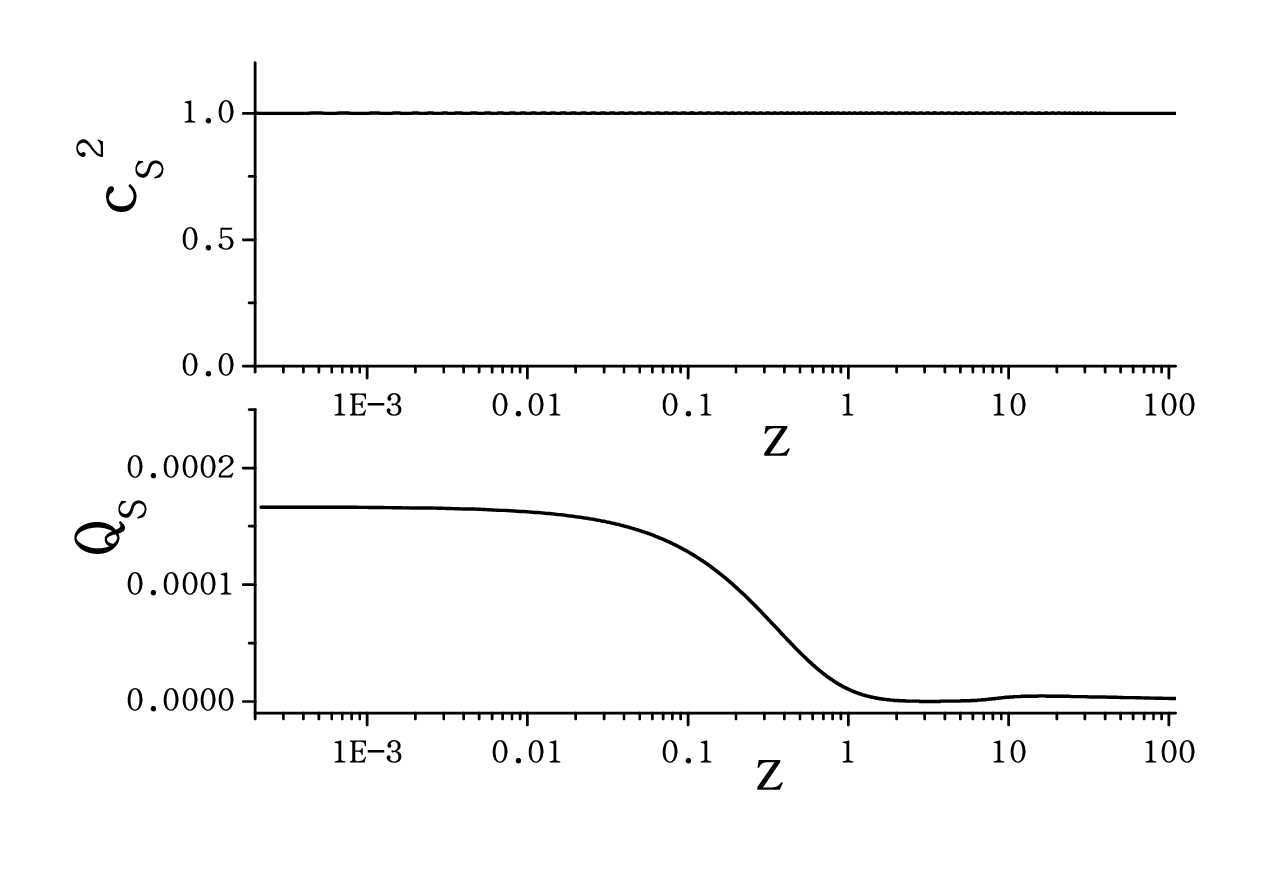}
	\caption{{\it{The sound speed square $c_{S}^{2}$ given in (\ref{cscon}) 
and the quantity $Q_{S}$    given in (\ref{Qscon}), as  functions of the 
redshift for Model I    with $V_0=0.08$    and with $\xi=1.3$ in  $H_0$  
units. The model is free from 
ghost and Laplacian instabilities. }}}
	\label{CSQS}
\end{figure}

Finally, we examine the stability of the obtained solution, by 
investigating the sound speed square $c_{S}^{2}$ given in (\ref{cscon}) and the 
quantity $Q_{S}$    given in (\ref{Qscon}). In Fig. 
\ref{CSQS} we depict their evolution as a function of the redshift for the 
background solution given above. As we can see, the stability conditions are 
always satisfied, and hence the obtained 
solutions are free from ghost and Laplacian instabilities (we mention that for 
a general $G_5\neq0$ the $c_{S}^{2}$ is not identically 1, however for the 
chosen 
$G_5(X)$ with the chosen $\xi$ and the imposed initial conditions we can bring 
it to be almost 1 during the whole evolution). Additionally,  the value of 
$Q_S$ for large $z$ is close to zero, however one can verify that it remains 
always  positive, and  for    $z\gg 1$  (i.e. for $z\rightarrow 1000$) 
it is around 0.005.

\begin{figure}[H]
%\hspace{-0.3cm}
	\includegraphics[width=0.45\textwidth]{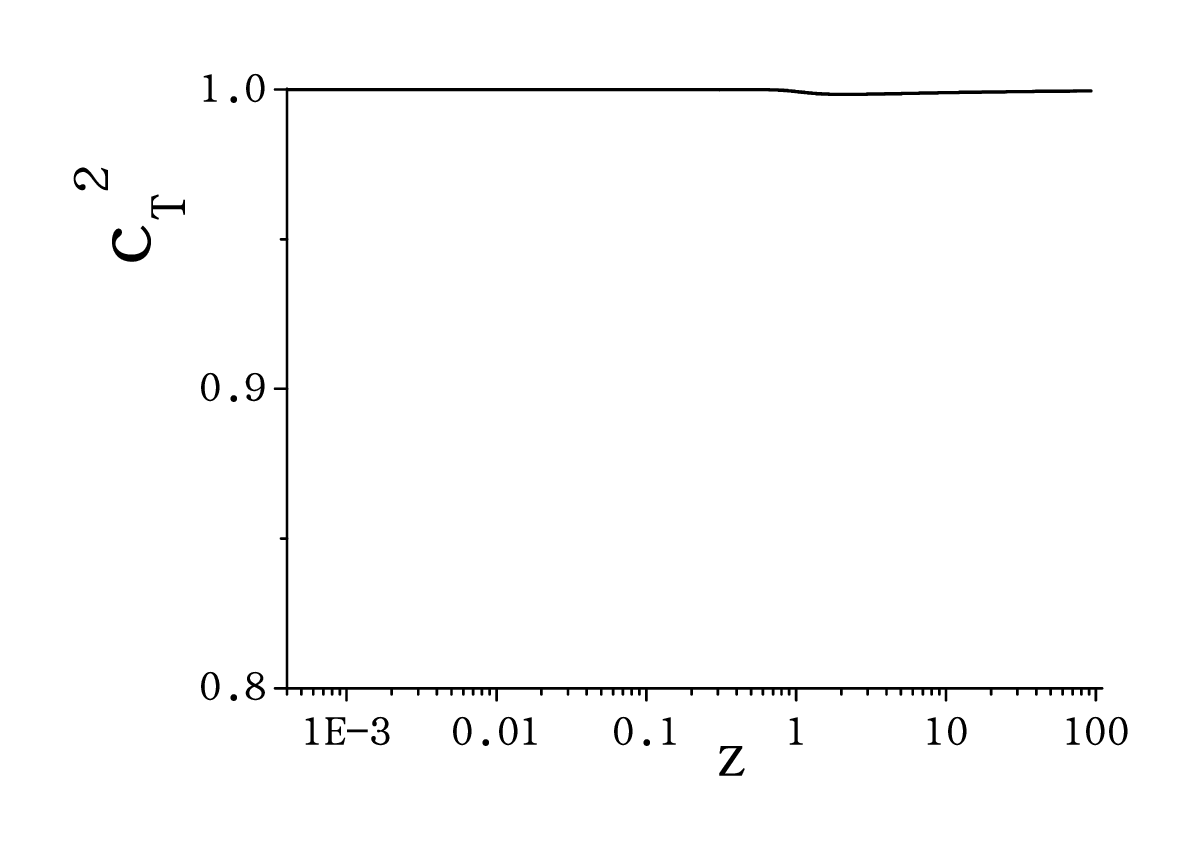}
	\caption{{\it{The gravitational-wave speed square $c_{T}^{2}$ given in 
(\ref{cTcon2})  as a function of the 
redshift, for Model I   with $V_0=0.08$  and with $\xi=1.3$ in  $H_0$  
units. 	 
}}}
	\label{CTFig}
\end{figure}

 Lastly, in   
Fig. 
\ref{CTFig} we depict the corresponding gravitational-wave speed square 
$c_{T}^{2}$ given in 
(\ref{cTcon2}). As we mentioned after (\ref{cTcon2}), 
in Horndeski theories the $G_5$ terms in general lead to a gravitational-wave 
speed different than $1$ (namely than the light speed). In 
the present  model, indeed the gravitational wave speed is not identically one, 
however for the chosen parameter value, since the 
$G_5$ terms are small  comparing to $G_4$, and moreover since $\ddot{\phi}$ is 
of the  same order of ${\dot{\phi}} H$, the gravitational-wave speed  is very 
close to 1. Specifically, the numerical difference between $c_T$ and unity for 
small $z$ ($z<0.5$) is less than  $\lesssim 10^{-15}$, while for $z>1$ it 
becomes larger, namely around $5\times 10^{-10}$. However, this is not in 
contradiction with  LIGO-VIRGO bounds \cite{Ezquiaga:2017ekz}, 
since both GW170817 and GW190425 neutron-star - neutron-star merger events are 
at very 
close distances, namely at redshifts around 0.01, and thus strictly speaking   
the observational verification of the gravitational-wave speed is only for very 
low redshifts (and also for the specific frequency range of LIGO-VIRGO). In 
summary, the present model is  able to pass the  LIGO-VIRGO bounds, however one 
could still try to  construct Horndeski models that can   alleviate the tension 
but have $c_T$ even more close to $1$.

\subsection{Model IΙ: $G_5(X)=\lambda X^{4}$}

The second model we consider is the one with  $G_5(X)=\lambda X^{4}$, 
i.e $G_5$ has a quartic   dependence on the field's kinetic energy. In this 
case (\ref{rhode}) and (\ref{pde})  respectively become
\begin{eqnarray}
\label{rhode2}
&&\!\!\!\!\!\!\!\!\!\!\!\!\!\!\!\!\rho_{DE}= 
\frac{\dot{\phi}^2}{2}+V_0\phi+\frac{11}{2}\lambda H^3 
\dot{\phi}^9 ,\\
\label{pde2}
&& \!\!\!\!\!\!\!\!\!\!\!\!\!\!\!\!p_{DE}= 
\frac{\dot{\phi}^2}{2}-V_0\phi-\frac{\lambda 
\dot{\phi}^8}{2}\left(
2H^3\dot{\phi}+2H\dot{H}\dot{\phi}+9H^2 \ddot{\phi}
\right).
 \end{eqnarray}

\begin{figure}[H]
\hspace{-0.3cm}
	\includegraphics[width=0.5\textwidth]{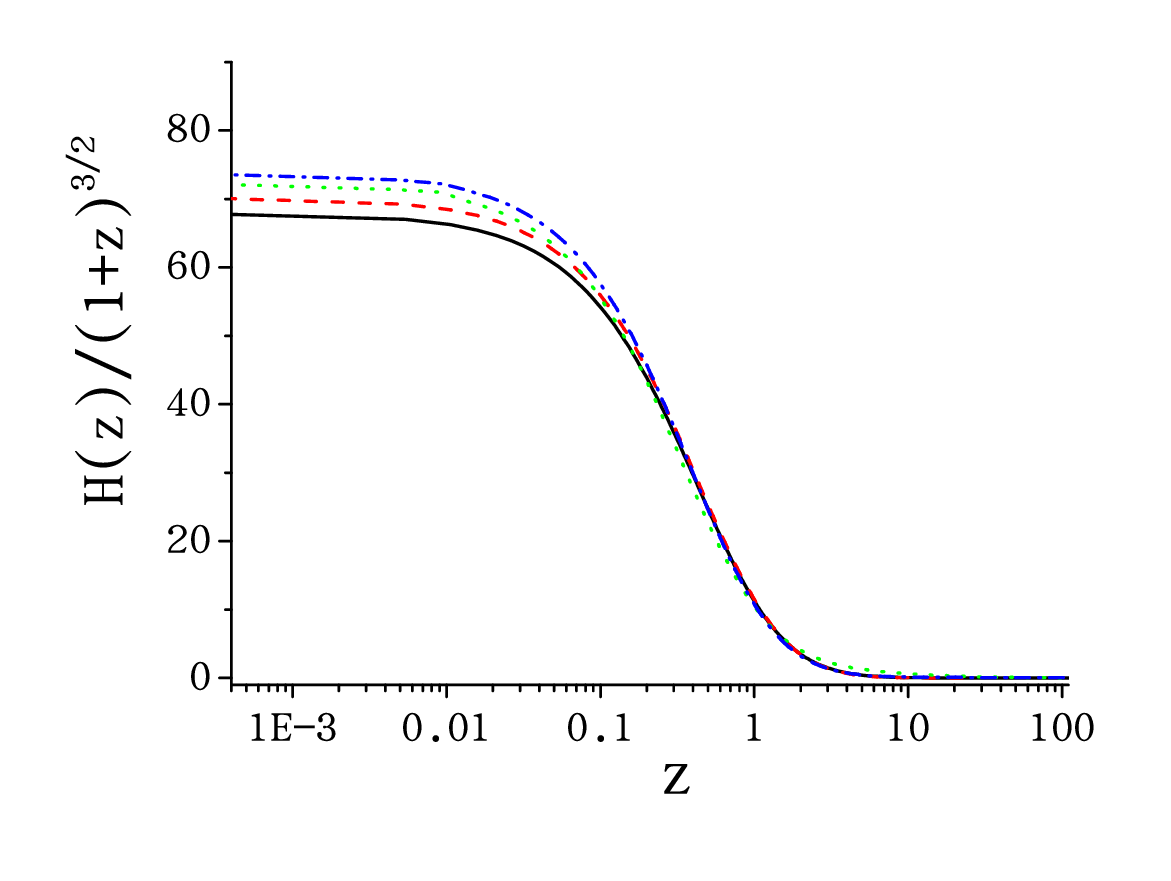} 
	\caption{{\it{
	The normalized $H(z)/(1+z)^{3/2}$ as a 
function of the redshift, for $\Lambda$CDM cosmology (black - solid) and for   
Model II   with $V_0=1$   and with $G_5(X)=\lambda X^4$, for 
$\lambda=0.5$ (red - dashed),  
$\lambda=0.9$ 
(green - 
dotted)
 and  $\lambda=1$ (blue - dashed-dotted), in  $H_0$  units. We have imposed 
$\Omega_{m0}\approx0.31$.}}}
	\label{lambdaX4}
\end{figure}

In Fig.  \ref{lambdaX4} we present   $H(z)/(1+z)^{3/2}$ as a 
function of the redshift for $\Lambda$CDM scenario and for our model with  
various choices of 
$\lambda$. The two models coincide at high and 
intermediate redshifts, but at small redshifts   Model II gives   
higher value, while  $H_0$ depends on the model 
parameter $\lambda$. Specifically, it can be around $H_0 \approx 74$ 
km/s/Mpc for $\lambda=1$ in  $H_0$  units,  and in general the tension can 
be alleviated at 3$\sigma$ if  
$0.5<\lambda<1.2$  (in Planck  units we have $V_0\sim
10^{-61}$ and $\lambda \sim 
10^{510}$, and since $\lambda$ has dimensions of $[M]^{-17}$ we acquire
$\lambda^{1/17}\sim 10^{30}$GeV$^{-1}$).
 Thus, we observe that   this  
kinetic-depedent sub-class of
Horndeski/generalized Galileon gravity can also alleviate the $H_0$ tension, 
since the  $G_5(X)$ term  that controls the friction term in the Friedmann 
equation is negligible at high redshifts, while it increases and plays a role 
at low redshifts. 
Lastly, we mention that 
the behavior of $c_{S}^{2}$ and $Q_{S}$ is similar to the one of Fig. 
\ref{CSQS}, i.e. the scenario at hand is free from ghost and 
Laplacian instabilities.

\begin{figure}[H]
	\centering
	\includegraphics[width=0.41\textwidth]{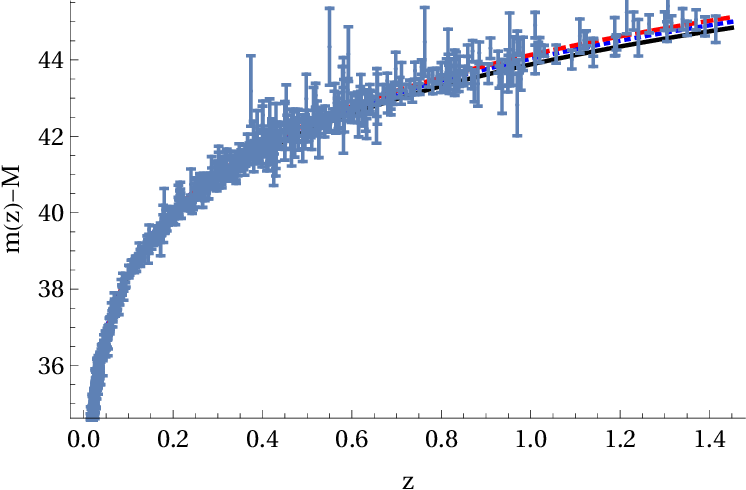}
	\caption{{\it{The apparent minus absolute magnitude 
	 predicted theoretically for Model I   with $V_0=0.08$    and with 
$\xi=1.3$ 
	 	  (red - dashed) and 
for Model II with $V_0=1$  and with $\lambda=1$  (blue - dotted)   in  
$H_0$  units,  on top of 
the Pantheon SNIa data points from
\cite{Scolnic:2017caz}.
For comparison  we depict the $\Lambda$CDM  curve  (black - solid) too.}}}
	\label{fdatafig}
\end{figure}

 We close this section by  confronting the models at hand with 
Supernovae 
type Ia (SNIa) and Cosmic Chronometer cosmological data. In particular, 
concerning SNIa it is known that 
 \begin{equation}
2.5 \log\left[\frac{L}{l(z)}\right] = \mu \equiv m(z) - M = 5 
\log\left[\frac{d_L(z)_{\text{obs}}}{Mpc}\right]  + 25~,
 \end{equation}
with $l(z)$ and  $m(z)$ the apparent luminosity and apparent 
magnitude, and  $L$ and $M$    the absolute luminosity and 
magnitude, respectively, while $d_L(z)_{\text{obs}}$ is the luminosity distance.
On the other hand, the theoretical value of the  
luminosity distance is
\begin{equation}
d_{L}\left(z\right)_\text{th}\equiv\left(1+z\right)
\int^{z}_{0}\frac{dz'}{H\left(z'\right)}~.
\end{equation}
 Since we know the evolution of  $H(z)$ in our models, as well as 
$H_{\Lambda\text{CDM}}(z)$, in Fig.   \ref{fdatafig} we 
depict the apparent minus absolute magnitude predicted theoretically  
for our models as well as   $\Lambda$CDM cosmology, on top of the  binned 
Pantheon sample   SNIa 
data points from
\cite{Scolnic:2017caz}. As we can see,   the 
agreement  is very good, and the proposed models have a slightly higher 
accelerating behavior, as expected.

\begin{figure}[H]
	\centering
	\includegraphics[width=0.41\textwidth]{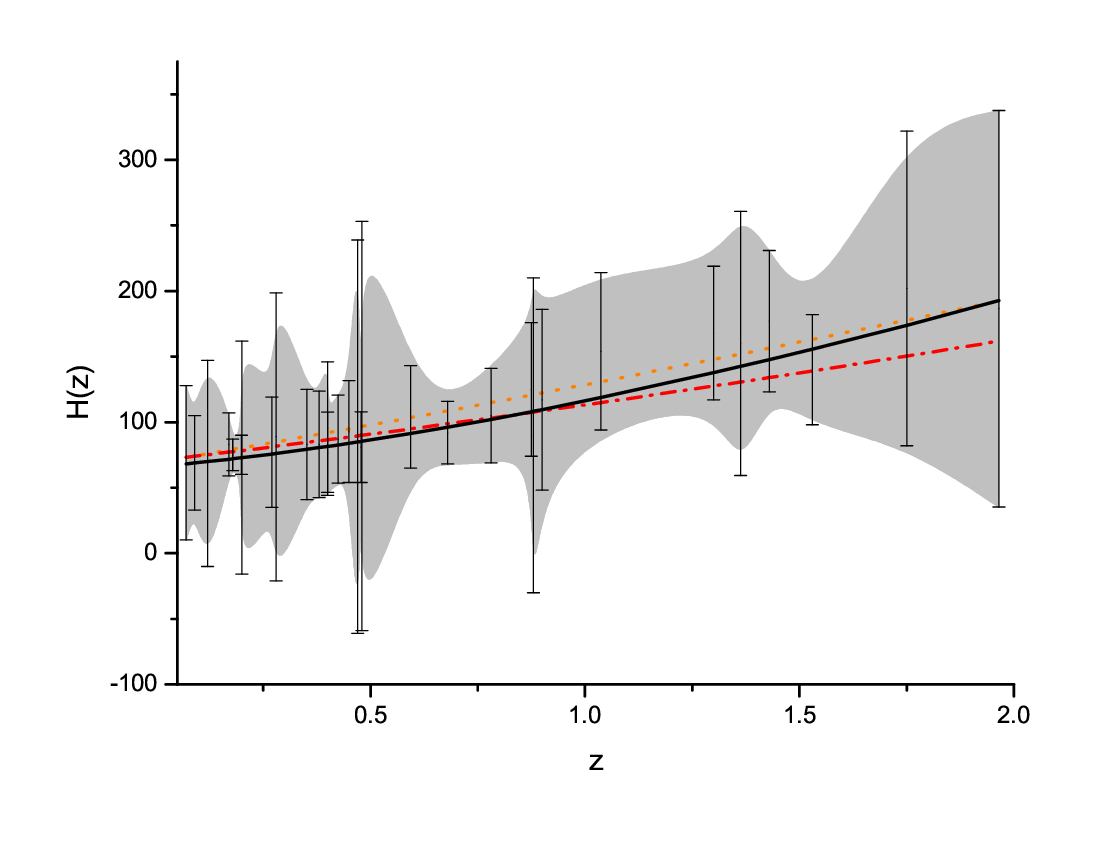}
	\caption{{\it{ The $H(z)$ in units of Km/s/Mpc as a 
				function of the redshift,
				for Model I   with $V_0=0.08$   and with $\xi=1.3$  
(red-dashed-dotted), 
				and 
				for Model II  with $V_0=1$   and with $\lambda=1$ 
(orange-dashed), in  $H_0$  units, on   top of the Cosmic Chronometers data 
points from \cite{Yu:2017iju} at $3\sigma$ confidence level. For comparison  we 
depict the $\Lambda$CDM  curve  (black - solid) too. We have imposed 
				$\Omega_{m0}\approx0.31$. }} }
	\label{fdata2fig}
\end{figure}

 Additionally, the Cosmic Chronometer (CC)
 datasets is based on the  measurements of  $H(z)$  using the relative ages of 
   massive and passively evolving galaxies and the corresponding estimation of 
 $dz/dt$   \cite{Jimenez:2001gg}. In Fig.   \ref{fdata2fig}   we   confront the 
theoretically predicted $H(z)$ behavior, as well as  the one of $\Lambda$CDM 
cosmology, with the $H(z)$ CC data from \cite{Yu:2017iju} at 
$3\sigma$ confidence level. The agreement is very good, and the $H(z)$ 
evolution of the proposed models lies within the prediction of the direct 
measurements of the $H(z)$ from the CC data, having again a 
slightly higher accelerating behavior at low redshifts, for the parameter 
sets $\left\lbrace \Omega_{m0},V_0,\xi\right\rbrace\,$  = $\left\lbrace 
0.31,\,0.08,\,1.3\right\rbrace\, $ and 
$\left\lbrace\Omega_{m0},V_0,\lambda\right\rbrace\, $ = 
$\left\lbrace0.31,\,1,\,1\right\rbrace$.

\!
 In summary,   there exist regions of the free parameters 
that are able to reproduce the observed Hubble function evolution and at late 
times potentially alleviate the $H_0$ tension, implying also the viability of 
the examined models. 
Definitely, in order to 
conclude on whether a specific model can alleviate the cosmological tensions, a 
full confrontation with all observational datasets is required.   The present 
work is just a first approach on the subject, in order 
to reveal the mechanism that is able to lift the present Hubble parameter value 
comparing to $\Lambda$CDM scenario (following the general requirements  of   
 \cite{Heisenberg:2022lob,Abdalla:2022yfr}).  
  The detailed verification of viability for the proposed 
models and their results, applying likelihood analysis and model selection 
criteria 
on full cosmological datasets, lies  beyond the scope of the current work and 
will be presented in a forthcoming project.

 \section{Conclusions}
 \label{Conclusions} 

The $H_0$ tension, unless it is caused by some unknown systematics or is related
with some basic data-handling error, may provide a strong indication towards 
the modification of Standard Model of cosmology. In the present work we 
investigated the possibility for its alleviation through Horndeski/generalized 
galileon gravity. 

In particular, knowing that the terms depending on $G_5$ control the friction 
term in the Friedmann equation, we constructed specific sub-classes depending 
only on the field's kinetic energy $X$. Since the kinetic energy is small at 
high redshifts, namely at redshifts which affected the CMB structure, the 
deviations from $\Lambda$CDM cosmology are negligible, however as time passes 
$X$ increases in a 
controlled way  and it leads to a decrease in the effective Newton's 
constant, and thus at low redshifts $H(z)$  acquires increased values.

We considered two Models, one with quadratic and one with quartic dependence on 
the  field's kinetic energy. In both cases we showed that at high and 
intermediate redshifts  the Hubble function behaves identically to that of  
$\Lambda$CDM scenario, however at low redshifts it acquires increased  values, 
resulting to  $H_0 \approx 74$ km/s/Mpc for particular parameter choices.
  Hence, these sub-classes of Horndeski/generalized Galileon gravity can 
alleviate 
the $H_0$ tension. 
We mention that the above behavior is obtained without a tuning in the initial 
conditions of  $\phi$ and $\dot\phi$ (we  do not have much freedom since we set 
$\Omega_{DE0}\approx0.7$  and moreover we desire to have  $w_{DE}(z=0)$ around 
$-1$), however the amount of tuning comes mainly in the selection of the 
functions   $G_i$'s and $ K(\phi,X)$, since only a small subclass of them can 
fulfill the above requirements.

As a self-consistency test we examined the behavior of scalar metric 
perturbations, showing that the conditions for absence of ghost and Laplacian 
instabilities are fulfilled throughout the evolution, and hence that the 
proposed solutions are stable and free from pathologies.  Finally, for 
completeness we confronted the proposed models with SNIa and Cosmic Chronometer 
data, as a first 
evidence that they are viable and in agreement with observations.

In summary, in this pilot project we showed that the  $H_0$ tension can be 
alleviated in  the 
modified gravity framework of Horndeski/generalized Galileon theory, due to 
the weakening of gravity at low redhifts by the terms depending on the 
scalar field's kinetic energy. Definitely, in order to obtain a more concrete 
verification of the above result one should perform a full observational 
confrontation, using the full   datasets, namely data from  SNIa, 
Baryonic Acoustic Oscillations (BAO), Cosmic Chronometers (CC),  Redshift 
Space Distortion (RSD), Cosmic Microwave Background (CMB) shift 
temperature and polarization, and f$\sigma_8$ observations, performing also 
the comparison to $\Lambda$CDM  concordance scenario using various information 
criteria. Such a full and detailed analysis, lies beyond the scope of 
this first work, and it is left for a future project.

\begin{acknowledgments} 
The authors would like to acknowledge the contribution of the COST Action 
CA18108 ``Quantum Gravity Phenomenology in the multi-messenger approach''.
\end{acknowledgments}


\begin{thebibliography}{99}
 
 %\cite{Sahni:1999gb}
\bibitem{Sahni:1999gb}
V.~Sahni and A.~A.~Starobinsky,
%``The Case for a positive cosmological Lambda term,''
Int.\ J.\ Mod.\ Phys. D \textbf{9}, 373-444 (2000)
[\href{\arxiv/astro-ph/9904398}{astro-ph/9904398}].


 %\cite{Peebles:2002gy}
\bibitem{Peebles:2002gy} 
  P.~J.~E.~Peebles and B.~Ratra,
  %The Cosmological constant and dark energy}},
   Rev.\ Mod.\ Phys.\  {\bf 75}, 559 (2003)
  [\href{\arxiv/astro-ph/0207347}{astro-ph/0207347}].
 
  
  
  
 \bibitem{Copeland:2006wr}
  E.~J.~Copeland, M.~Sami and S.~Tsujikawa,
 %Dynamics of dark energy}},
  Int.\ J.\ Mod.\ Phys.\  D {\bf 15}, 1753 (2006)
 [\href{http://xxx.lanl.gov/abs/hep-th/0603057}
{\tt arXiv:hep-th/0603057}].
 

 
 %\cite{Cai:2009zp}
\bibitem{Cai:2009zp}
  Y.~-F.~Cai, E.~N.~Saridakis, M.~R.~Setare and J.~-Q.~Xia,
  %Quintom Cosmology: Theoretical implications and observations}},
  Phys.\ Rept.\  {\bf 493}, 1 (2010)  [\href{http://xxx.lanl.gov/abs/0909.2776}
{{\tt arXiv:0909.2776}}].




%\cite{CANTATA:2021ktz}
\bibitem{CANTATA:2021ktz}
E.~N.~Saridakis \textit{et al.} [CANTATA],
%``Modified Gravity and Cosmology: An Update by the CANTATA Network,''
[\href{\arxiv/arXiv:2105.12582}{arXiv:2105.12582}].
%25 citations counted in INSPIRE as of 18 Aug 2021
   
 
 

%\cite{Capozziello:2011et}
\bibitem{Capozziello:2011et}
  S.~Capozziello and M.~De Laurentis,
%Extended Theories of Gravity}},
  Phys.\ Rept.\  {\bf 509}, 167 (2011)
[\href{http://xxx.lanl.gov/abs/1108.6266}
{{\tt arXiv:1108.6266}}].
 

%\cite{Cai:2015emx}
\bibitem{Cai:2015emx} 
  Y.~F.~Cai, S.~Capozziello, M.~De Laurentis and E.~N.~Saridakis,
 %f(T) teleparallel gravity and cosmology}},
  Rept.\ Prog.\ Phys.\  {\bf 79}, 106901 (2016).
 [\href{http://xxx.lanl.gov/abs/1511.07586}
 {arXiv:1511.07586 [gr-qc]}].
 
  
 
%\cite{Aghanim:2018eyx}
\bibitem{Aghanim:2018eyx} 
 N.~Aghanim {\it et al.} [Planck Collaboration],
% Planck 2018 results. VI. Cosmological parameters}},(2018)
  [\href{\arxiv/arXiv:1807.06209}{arXiv:1807.06209}].
 

%\cite{Riess:2019cxk}
\bibitem{Riess:2019cxk} 
  A.~G.~Riess, S.~Casertano, W.~Yuan, L.~M.~Macri and D.~Scolnic,
% Large Magellanic Cloud Cepheid Standards Provide a 1\% Foundation for the 
%Determination of the Hubble Constant and Stronger Evidence for Physics beyond 
%$\Lambda$CDM}},
  Astrophys.\ J.\  {\bf 876}, 85 (2019)
  %doi:10.3847/1538-4357/ab1422
  [\href{\arxiv/arXiv:1903.07603}{arXiv:1903.07603}].
  

%\cite{Wong:2019kwg}
\bibitem{Wong:2019kwg} 
  K.~C.~Wong {\it et al.},
 %H0LiCOW XIII. A 2.4\% measurement of $H_{0}$ from lensed quasars: 
%$5.3\sigma$ tension between early and late-Universe probes}},
   [\href{\arxiv/arXiv:1907.04869}{arXiv:1907.04869}].
  

 

%\cite{Alam:2016hwk}
\bibitem{Alam:2016hwk} 
  S.~Alam {\it et al.} [BOSS Collaboration],
 %The clustering of galaxies in the completed SDSS-III Baryon Oscillation 
%Spectroscopic Survey: cosmological analysis of the DR12 galaxy sample}},
  Mon.\ Not.\ Roy.\ Astron.\ Soc.\  {\bf 470}, 2617 (2017)
  %doi:10.1093/mnras/stx721
  [\href{\arxiv/arXiv:1607.03155}{arXiv:1607.03155 }].
  

%\cite{Ata:2017dya}
\bibitem{Ata:2017dya} 
  M.~Ata {\it et al.},
  %The clustering of the SDSS-IV extended Baryon Oscillation Spectroscopic 
%Survey DR14 quasar sample: first measurement of baryon acoustic oscillations 
%between redshift 0.8 and 2.2}},
  Mon.\ Not.\ Roy.\ Astron.\ Soc.\  {\bf 473}, 4773 (2018)
  %doi:10.1093/mnras/stx2630
  [\href{\arxiv/arXiv:1705.06373}{arXiv:1705.06373}].
  %%CITATION = doi:10.1093/mnras/stx2630;%%
  
  %\cite{Zarrouk:2018vwy}
  \bibitem{Zarrouk:2018vwy} 
  P.~Zarrouk {\it et al.},
  %The clustering of the SDSS-IV extended Baryon Oscillation Spectroscopic 
 % 		Survey DR14 quasar sample: measurement of the growth rate of structure 
%from the 
 % 		anisotropic correlation function between redshift 0.8 and 2.2}},
  Mon.\ Not.\ Roy.\ Astron.\ Soc.\  {\bf 477},  1639 (2018)
  %doi:10.1093/mnras/sty506
  [\href{\arxiv/arXiv:1801.03062}{arXiv:1801.03062}].
  
  

%\cite{DiValentino:2021izs}
\bibitem{DiValentino:2021izs}
E.~Di Valentino, O.~Mena, S.~Pan, L.~Visinelli, W.~Yang, A.~Melchiorri, 
D.~F.~Mota, A.~G.~Riess and J.~Silk,
%``In the realm of the Hubble tension\textemdash{}a review of solutions,''
Class. Quant. Grav. \textbf{38}, no.15, 153001 (2021)
[\href{\arxiv/arXiv:2103.01183}{arXiv:2103.01183}].
%175 citations counted in INSPIRE as of 24 Sep 2021

%\cite{DiValentino:2020zio}
\bibitem{DiValentino:2020zio}
E.~Di Valentino, L.~A.~Anchordoqui, O.~Akarsu, Y.~Ali-Haimoud, L.~Amendola, 
N.~Arendse, M.~Asgari, M.~Ballardini, S.~Basilakos and E.~Battistelli, 
\textit{et al.}
%Cosmology Intertwined II: The Hubble Constant Tension}},
[\href{\arxiv/arXiv:2008.11284}{arXiv:2008.11284}].

%\cite{DiValentino:2015ola}
\bibitem{DiValentino:2015ola}
E.~Di Valentino, A.~Melchiorri and J.~Silk,
%``Beyond six parameters: extending $\Lambda$CDM,''
Phys. Rev. D \textbf{92}, no.12, 121302 (2015)
[\href{\arxiv/arXiv:1507.06646}{arXiv:1507.06646}].

%\cite{Bernal:2016gxb}
\bibitem{Bernal:2016gxb}
J.~L.~Bernal, L.~Verde and A.~G.~Riess,
%``The trouble with $H_0$,''
JCAP \textbf{10}, 019 (2016)
 \href{\arxiv/arXiv:1607.05617}{arXiv:1607.05617}].
%445 citations counted in INSPIRE as of 21 Sep 2021




%\cite{Pan:2019gop}
\bibitem{Pan:2019gop}
S.~Pan, W.~Yang, E.~Di Valentino, E.~N.~Saridakis and S.~Chakraborty,
%Interacting scenarios with dynamical dark energy: Observational 
%constraints and alleviation of the $H_0$ tension}},
Phys. Rev. D \textbf{100}, 103520 (2019)
[\href{\arxiv/arXiv:1907.07540}{arXiv:1907.07540}].


%\cite{Pan:2019jqh}
\bibitem{Pan:2019jqh}
S.~Pan, W.~Yang, C.~Singha and E.~N.~Saridakis,
%Observational constraints on sign-changeable interaction models and 
%alleviation of the $H_0$ tension}},
Phys. Rev. D \textbf{100}, 083539 (2019)
[\href{\arxiv/arXiv:1903.10969}{arXiv:1903.10969}].





%\cite{Yang:2018prh}
\bibitem{Yang:2018prh}
W.~Yang, S.~Pan, E.~Di Valentino and E.~N.~Saridakis,
%Observational constraints on dynamical dark energy with pivoting 
%redshift}},
Universe \textbf{5}, 219 (2019)
[\href{\arxiv/arXiv:1811.06932}{arXiv:1811.06932}].



%\cite{Yang:2018qmz}
\bibitem{Yang:2018qmz}
W.~Yang, S.~Pan, E.~Di Valentino, E.~N.~Saridakis and S.~Chakraborty,
%Observational constraints on one-parameter dynamical dark-energy 
%parametrizations and the $H_0$ tension}},
Phys. Rev. D \textbf{99}, 043543 (2019)
[\href{\arxiv/arXiv:1810.05141}{arXiv:1810.05141}].



 %\cite{Kumar:2016zpg}
 \bibitem{Kumar:2016zpg}
 S.~Kumar and R.~C.~Nunes,
 %Probing the interaction between dark matter and dark energy in the 
%presence of massive neutrinos}},
 Phys. Rev. D \textbf{94}, no.12, 123511 (2016)
 [\href{\arxiv/arXiv:1608.02454}{arXiv:1608.02454}][astro-ph.CO].
 %132 citations counted in INSPIRE as of 22 Sep 2021 

%\cite{DiValentino:2017iww}
\bibitem{DiValentino:2017iww}
E.~Di Valentino, A.~Melchiorri and O.~Mena,
Phys. Rev. D \textbf{96}, no.4, 043503 (2017)
[\href{\arxiv/arXiv:1704.08342}{arXiv:1704.08342}][astro-ph.CO].
%244 citations counted in INSPIRE as of 22 Sep 2021

%\cite{DiValentino:2017oaw}
\bibitem{DiValentino:2017oaw}
E.~Di Valentino, C.~B\o{}ehm, E.~Hivon and F.~R.~Bouchet,
%Reducing the $H_0$ and $\sigma_8$ tensions with Dark Matter-neutrino 
%interactions}},
Phys. Rev. D \textbf{97}, no.4, 043513 (2018)
[\href{\arxiv/arXiv:1710.02559}{arXiv:1710.02559}] .
%110 citations counted in INSPIRE as of 22 Sep 2021

%\cite{Binder:2017lkj}
\bibitem{Binder:2017lkj}
T.~Binder, M.~Gustafsson, A.~Kamada, S.~M.~R.~Sandner and M.~Wiesner,
%Reannihilation of self-interacting dark matter}},
Phys. Rev. D \textbf{97}, no.12, 123004 (2018)
[\href{\arxiv/arXiv:1712.01246}{arXiv:1712.01246}] [astro-ph.CO].
%38 citations counted in INSPIRE as of 22 Sep 2021

%\cite{DiValentino:2017zyq}
\bibitem{DiValentino:2017zyq}
E.~Di Valentino, A.~Melchiorri, E.~V.~Linder and J.~Silk,
%Constraining Dark Energy Dynamics in Extended Parameter Space}},
Phys. Rev. D \textbf{96}, no.2, 023523 (2017)
[\href{\arxiv/arXiv:1704.00762}{arXiv:1704.00762}] .
%150 citations counted in INSPIRE as of 22 Sep 2021
%\cite{Sola:2017znb}

\bibitem{Sola:2017znb}
J.~Sol\`a, A.~G\'omez-Valent and J.~de Cruz P\'erez,
%``The $H_0$ tension in light of vacuum dynamics in the Universe,''
Phys. Lett. B \textbf{774}, 317-324 (2017)
[\href{\arxiv/arXiv:1705.06723}{arXiv:1705.06723}] [astro-ph.CO].
%109 citations counted in INSPIRE as of 22 Sep 2021

%\cite{Yang:2018euj}
\bibitem{Yang:2018euj}
W.~Yang, S.~Pan, E.~Di Valentino, R.~C.~Nunes, S.~Vagnozzi and D.~F.~Mota,
%Tale of stable interacting dark energy, observational signatures, and the 
%$H_0$ tension}},
JCAP \textbf{09}, 019 (2018)
[\href{\arxiv/arXiv:1805.08252}{arXiv:1805.08252}].
%174 citations counted in INSPIRE as of 22 Sep 2021

%\cite{DEramo:2018vss}
\bibitem{DEramo:2018vss}
F.~D'Eramo, R.~Z.~Ferreira, A.~Notari and J.~L.~Bernal,
%``Hot Axions and the $H_0$ tension,''
JCAP \textbf{11}, 014 (2018)
[\href{\arxiv/arXiv:1808.07430}{arXiv:1808.07430}].
%111 citations counted in INSPIRE as of 22 Sep 2021

%\cite{Poulin:2018cxd}
\bibitem{Poulin:2018cxd}
V.~Poulin, T.~L.~Smith, T.~Karwal and M.~Kamionkowski,
%Early Dark Energy Can Resolve The Hubble Tension}},
Phys. Rev. Lett. \textbf{122}, no.22, 221301 (2019)
[\href{\arxiv/arXiv:1811.04083}{arXiv:1811.04083}].
%336 citations counted in INSPIRE as of 22 Sep 2021

%\cite{Shafieloo:2016bpk}
\bibitem{Shafieloo:2016bpk}
A.~Shafieloo, D.~K.~Hazra, V.~Sahni and A.~A.~Starobinsky,
%``Metastable Dark Energy with Radioactive-like Decay,''
Mon. Not. Roy. Astron. Soc. \textbf{473}, no.2, 2760-2770 (2018)
[\href{\arxiv/arXiv:1610.05192}{arXiv:1610.05192}] [astro-ph.CO].
%26 citations counted in INSPIRE as of 22 Sep 2021


%\cite{Lancaster:2017ksf}
\bibitem{Lancaster:2017ksf}
L.~Lancaster, F.~Y.~Cyr-Racine, L.~Knox and Z.~Pan,
%A tale of two modes: Neutrino free-streaming in the early universe}},
JCAP \textbf{07}, 033 (2017)
[\href{\arxiv/arXiv:1704.06657}{arXiv:1704.06657}].
%78 citations counted in INSPIRE as of 22 Sep 2021


%\cite{Berghaus:2019cls}
\bibitem{Berghaus:2019cls}
K.~V.~Berghaus and T.~Karwal,
%Thermal Friction as a Solution to the Hubble Tension}},
Phys. Rev. D \textbf{101}, no.8, 083537 (2020)
[\href{\arxiv/arXiv:1911.06281}{arXiv:1911.06281}] [astro-ph.CO].
%47 citations counted in INSPIRE as of 22 Sep 2021


%\cite{Pandey:2019plg}
\bibitem{Pandey:2019plg}
K.~L.~Pandey, T.~Karwal and S.~Das,
%Alleviating the $H_0$ and $\sigma_8$ anomalies with a decaying dark matter 
%model}},
JCAP \textbf{07}, 026 (2020)
[\href{\arxiv/arXiv:1902.10636}{arXiv:1902.10636}].
%87 citations counted in INSPIRE as of 22 Sep 2021

%\cite{Adhikari:2019fvb}
\bibitem{Adhikari:2019fvb}
S.~Adhikari and D.~Huterer,
%Super-CMB fluctuations and the Hubble tension}},
Phys. Dark Univ. \textbf{28}, 100539 (2020)
[\href{\arxiv/arXiv:1905.02278}{arXiv:1905.02278}].
%29 citations counted in INSPIRE as of 22 Sep 2021



%\cite{Benisty:2019pxb}
\bibitem{Benisty:2019pxb}
D.~Benisty,
%``Decaying coupled Fermions to curvature and the $H_0$ tension,''
[\href{\arxiv/arXiv:1912.11124}{arXiv:1912.11124}].
 

%\cite{Perez:2020cwa}
\bibitem{Perez:2020cwa}
A.~Perez, D.~Sudarsky and E.~Wilson-Ewing,
%``Resolving the $H_0$ tension with diffusion,''
Gen. Rel. Grav. \textbf{53}, no.1, 7 (2021)
%doi:10.1007/s10714-020-02781-0
[\href{\arxiv/arXiv:2001.07536}{arXiv:2001.07536}].
%16 citations counted in INSPIRE as of 22 Sep 2021

%\cite{Pan:2020bur}
\bibitem{Pan:2020bur}
S.~Pan, W.~Yang and A.~Paliathanasis,
%Non-linear interacting cosmological models after Planck 2018 legacy release 
%and the $H_0$ tension}},
Mon. Not. Roy. Astron. Soc. \textbf{493}, no.3, 3114-3131 (2020)
%doi:10.1093/mnras/staa213
[\href{\arxiv/arXiv:2002.03408}{arXiv:2002.03408}].
%32 citations counted in INSPIRE as of 22 Sep 2021

%\cite{Benevento:2020fev}
\bibitem{Benevento:2020fev}
G.~Benevento, W.~Hu and M.~Raveri,
%Can Late Dark Energy Transitions Raise the Hubble constant?}},
Phys. Rev. D \textbf{101}, no.10, 103517 (2020)
[\href{\arxiv/arXiv:2002.11707}{arXiv:2002.11707}].
%71 citations counted in INSPIRE as of 22 Sep 2021



%\cite{Banerjee:2020xcn}
\bibitem{Banerjee:2020xcn}
A.~Banerjee, H.~Cai, L.~Heisenberg, E.~\'O.~Colg\'ain, M.~M.~Sheikh-Jabbari and 
T.~Yang,
%``Hubble sinks in the low-redshift swampland,''
Phys. Rev. D \textbf{103}, no.8, L081305 (2021)
%doi:10.1103/PhysRevD.103.L081305
[\href{\arxiv/arXiv:2006.00244}{arXiv:2006.00244}].
 



%\cite{Elizalde:2020mfs}
\bibitem{Elizalde:2020mfs}
E.~Elizalde, M.~Khurshudyan, S.~D.~Odintsov and R.~Myrzakulov,
%``Analysis of the $H_0$ tension problem in the Universe with viscous dark 
%fluid,''
Phys. Rev. D \textbf{102}, no.12, 123501 (2020)
[\href{\arxiv/arXiv:2006.01879}{arXiv:2006.01879}].


 


%\cite{Alvarez:2020xmk}
\bibitem{Alvarez:2020xmk}
P.~D.~Alvarez, B.~Koch, C.~Laporte and \'A.~Rinc\'on,
%Can scale-dependent cosmology alleviate the $H_0$ tension?}},
JCAP \textbf{06}, 019 (2021)
[\href{\arxiv/arXiv:2009.02311}{arXiv:2009.02311}].
%7 citations counted in INSPIRE as of 22 Sep 2021

%\cite{DeFelice:2020cpt}
\bibitem{DeFelice:2020cpt}
A.~De Felice, S.~Mukohyama and M.~C.~Pookkillath,
%Addressing $H_0$ tension by means of VCDM}},
Phys. Lett. B \textbf{816}, 136201 (2021)
[\href{\arxiv/arXiv:2009.08718}{arXiv:2009.08718}].
%12 citations counted in INSPIRE as of 22 Sep 2021

%\cite{Haridasu:2020pms}
\bibitem{Haridasu:2020pms}
B.~S.~Haridasu, M.~Viel and N.~Vittorio,
%``Sources of $H_0$-tension in dark energy scenarios,''
Phys. Rev. D \textbf{103}, no.6, 063539 (2021)
%doi:10.1103/PhysRevD.103.063539
[\href{\arxiv/arXiv:2012.10324}{arXiv:2012.10324}].


%\cite{Seto:2021xua}
\bibitem{Seto:2021xua}
O.~Seto and Y.~Toda,
%``Comparing early dark energy and extra radiation solutions to the Hubble 
%tension with BBN,''
Phys. Rev. D \textbf{103}, no.12, 123501 (2021)
%doi:10.1103/PhysRevD.103.123501
[\href{\arxiv/arXiv:2101.03740}{arXiv:2101.03740}].
%14 citations counted in INSPIRE as of 24 Sep 2021


%\cite{Bernal:2021yli}
\bibitem{Bernal:2021yli}
J.~L.~Bernal, L.~Verde, R.~Jimenez, M.~Kamionkowski, D.~Valcin and 
B.~D.~Wandelt,
%``The trouble beyond $H_0$ and the new cosmic triangles,''
Phys. Rev. D \textbf{103}, no.10, 103533 (2021)
%doi:10.1103/PhysRevD.103.103533
[\href{\arxiv/arXiv:2102.05066}{arXiv:2102.05066}].


%\cite{Alestas:2021xes}
\bibitem{Alestas:2021xes}
G.~Alestas and L.~Perivolaropoulos,
%``Late-time approaches to the Hubble tension deforming H(z), worsen the growth 
%tension,''
Mon. Not. Roy. Astron. Soc. \textbf{504}, no.3, 3956-3962 (2021)
%doi:10.1093/mnras/stab1070
[\href{\arxiv/arXiv:2103.04045}{arXiv:2103.04045}].
%13 citations counted in INSPIRE as of 24 Sep 2021

%\cite{Elizalde:2021kmo}
\bibitem{Elizalde:2021kmo}
E.~Elizalde, J.~Gluza and M.~Khurshudyan,
%``An approach to cold dark matter deviation and the $H_{0}$ tension problem by 
%using machine learning,''
[\href{\arxiv/arXiv:2104.01077}{arXiv:2104.01077}]

 %\cite{Krishnan:2021dyb}
\bibitem{Krishnan:2021dyb}
C.~Krishnan, R.~Mohayaee, E.~\'O.~Colg\'ain, M.~M.~Sheikh-Jabbari and L.~Yin,
%``Does Hubble tension signal a breakdown in FLRW cosmology?,''
Class. Quant. Grav. \textbf{38}, no.18, 184001 (2021)
 [\href{\arxiv/arXiv:2105.09790}{arXiv:2105.09790}]

 



%\cite{Theodoropoulos:2021hkk}
\bibitem{Theodoropoulos:2021hkk}
A.~Theodoropoulos and L.~Perivolaropoulos,
%``The Hubble Tension, the M Crisis of Late Time H(z) Deformation Models and 
the 
%Reconstruction of Quintessence Lagrangians,''
Universe \textbf{7}, no.8, 300 (2021)
%doi:10.3390/universe7080300
[\href{\arxiv/arXiv:2109.06256}{arXiv:2109.06256}].



%\cite{Anagnostopoulos:2019miu}
\bibitem{Anagnostopoulos:2019miu}
F.~K.~Anagnostopoulos, S.~Basilakos and E.~N.~Saridakis,
%Bayesian analysis of $f(T)$ gravity using $f\sigma_8$ data}},
Phys. Rev. D \textbf{100}, 083517 (2019)
[\href{\arxiv/arXiv:1907.07533}{arXiv:1907.07533}].

%\cite{El-Zant:2018bsc}
\bibitem{El-Zant:2018bsc}
A.~El-Zant, W.~El Hanafy and S.~Elgammal,
%$H_0$ Tension and the Phantom Regime: A Case Study in Terms of an Infrared 
%$f(T)$ Gravity}},
Astrophys. J. \textbf{871}, 210 (2019)
[\href{\arxiv/arXiv:1809.09390}{arXiv:1809.09390}].



%\cite{Braglia:2020auw}
\bibitem{Braglia:2020auw}
M.~Braglia, M.~Ballardini, F.~Finelli and K.~Koyama,
%Early modified gravity in light of the $H_0$ tension and LSS data}},
[\href{\arxiv/arXiv:2011.12934}{arXiv:2011.12934 }].



%\cite{Abadi:2020hbr}
\bibitem{Abadi:2020hbr}
T.~Abadi and E.~D.~Kovetz,
%Can Conformally Invariant Modified Gravity Solve The Hubble Tension?}},
[\href{\arxiv/arXiv:2011.13853}{arXiv:2011.13853} ].




%\cite{Cai:2019bdh}
\bibitem{Cai:2019bdh}
Y.~F.~Cai, M.~Khurshudyan and E.~N.~Saridakis,
%Model-independent reconstruction of $f(T)$ gravity from Gaussian 
%Processes}},
Astrophys. J. \textbf{888}, 62 (2020)
[\href{\arxiv/arXiv:1907.10813}{arXiv:1907.10813}].



%\cite{Escamilla-Rivera:2019ulu}
\bibitem{Escamilla-Rivera:2019ulu}
C.~Escamilla-Rivera and J.~Levi Said,
%Cosmological viable models in $f(T,B)$ theory as solutions to the $H_0$ 
%tension}},
Class. Quant. Grav. \textbf{37}, 165002 (2020)
[\href{\arxiv/arXiv:1909.10328}{arXiv:1909.10328}].


%\cite{Barker:2020gcp}
\bibitem{Barker:2020gcp}
W.~E.~V.~Barker, A.~N.~Lasenby, M.~P.~Hobson and W.~J.~Handley,
%Systematic study of background cosmology in unitary Poincar\'e gauge 
%theories with application to emergent dark radiation and $H_0$ tension}},
Phys. Rev. D \textbf{102}, 024048 (2020)
[\href{\arxiv/arXiv:2003.02690}{arXiv:2003.02690}].



%\cite{Wang:2020zfv}
\bibitem{Wang:2020zfv}
D.~Wang and D.~Mota,
%Can $f(T)$ gravity resolve the $H_0$ tension?}},
Phys. Rev. D \textbf{102}, 063530 (2020)
[\href{\arxiv/arXiv:2003.10095}{arXiv:2003.10095}].


%\cite{Ballardini:2020iws}
\bibitem{Ballardini:2020iws}
M.~Ballardini, M.~Braglia, F.~Finelli, D.~Paoletti, A.~A.~Starobinsky and 
C.~Umilt\`a,
%Scalar-tensor theories of gravity, neutrino physics, and the $H_0$ 
%tension}},
JCAP \textbf{10}, 044 (2020)
[\href{\arxiv/arXiv:2004.14349}{arXiv:2004.14349}].


%\cite{LinaresCedeno:2020uxx}
\bibitem{LinaresCedeno:2020uxx}
F.~X.~Linares Cede\~no and U.~Nucamendi,
%Revisiting cosmological diffusion models in Unimodular Gravity and the 
%$H_0$ tension}}, (2020)
[\href{\arxiv/arXiv:2009.10268}{arXiv:2009.10268}].



%\cite{Basilakos:2018arq}
\bibitem{Basilakos:2018arq}
S.~Basilakos, S.~Nesseris, F.~K.~Anagnostopoulos and E.~N.~Saridakis,
%Updated constraints on $f(T)$ models using direct and indirect 
%measurements 
%of the Hubble parameter}},
JCAP \textbf{08}, 008 (2018)
[\href{\arxiv/arXiv:1803.09278}{arXiv:1803.09278}].


%\cite{Adil:2021zxp}
\bibitem{Adil:2021zxp}
S.~A.~Adil, M.~R.~Gangopadhyay, M.~Sami and M.~K.~Sharma,
%``Late time acceleration due to generic modification of gravity and Hubble 
%tension,''
[\href{\arxiv/arXiv:2106.03093}{arXiv:2106.03093}].
%1 citations counted in INSPIRE as of 24 Sep 2021

%\cite{Odintsov:2020qzd}
\bibitem{Odintsov:2020qzd}
S.~D.~Odintsov, D.~S.~C.~G\'omez and G.~S.~Sharov,
%Analyzing the $H_0$ tension in $F(R)$ gravity models}}, (2020)
Nucl. Phys. B \textbf{966}, 115377 (2021)
[\href{\arxiv/arXiv:2011.03957}{arXiv:2011.03957}].

 



%\cite{DiValentino:2019jae}
\bibitem{DiValentino:2019jae}
E.~Di Valentino, A.~Melchiorri, O.~Mena and S.~Vagnozzi,
%Nonminimal dark sector physics and cosmological tensions}},
Phys. Rev. D \textbf{101}, 063502 (2020)
[\href{\arxiv/arXiv:1910.09853}{arXiv:1910.09853}].

 
 
 %\cite{Vagnozzi:2019ezj}
\bibitem{Vagnozzi:2019ezj}
S.~Vagnozzi,
%New physics in light of the $H_0$ tension: An alternative view}},
Phys. Rev. D \textbf{102}, 023518 (2020)
[\href{\arxiv/arXiv:1907.07569}{arXiv:1907.07569}].


%\cite{Hu:2015rva}
\bibitem{Hu:2015rva}
B.~Hu and M.~Raveri,
%Can modified gravity models reconcile the tension between the CMB 
%anisotropy and lensing maps in Planck-like observations?}},
Phys. Rev. D \textbf{91}, no.12, 123515 (2015)
[\href{\arxiv/arXiv:1502.06599}{arXiv:1502.06599}].
%19 citations counted in INSPIRE as of 22 Sep 2021

%\cite{Khosravi:2017hfi}
\bibitem{Khosravi:2017hfi}
N.~Khosravi, S.~Baghram, N.~Afshordi and N.~Altamirano,
%$H_0$ tension as a hint for a transition in gravitational theory}},
Phys. Rev. D \textbf{99}, no.10, 103526 (2019)
[\href{\arxiv/arXiv:1710.09366}{arXiv:1710.09366}].
%56 citations counted in INSPIRE as of 22 Sep 2021

%\cite{Belgacem:2017cqo}
\bibitem{Belgacem:2017cqo}
E.~Belgacem, Y.~Dirian, S.~Foffa and M.~Maggiore,
%``Nonlocal gravity. Conceptual aspects and cosmological predictions,''
JCAP \textbf{03}, 002 (2018)
[\href{\arxiv/arXiv:1712.07066}{arXiv:1712.07066}] [hep-th].
%107 citations counted in INSPIRE as of 22 Sep 2021

%\cite{Nunes:2018xbm}
\bibitem{Nunes:2018xbm}
R.~C.~Nunes,
%Structure formation in $f(T)$ gravity and a solution for $H_0$ tension}},
JCAP \textbf{05}, 052 (2018)
[\href{\arxiv/arXiv:1802.02281}{arXiv:1802.02281}].
%109 citations counted in INSPIRE as of 22 Sep 2021

%\cite{Lin:2018nxe}
\bibitem{Lin:2018nxe}
M.~X.~Lin, M.~Raveri and W.~Hu,
%``Phenomenology of Modified Gravity at Recombination,''
Phys. Rev. D \textbf{99}, no.4, 043514 (2019)
[\href{\arxiv/arXiv:1810.02333}{arXiv:1810.02333}].
%40 citations counted in INSPIRE as of 22 Sep 2021


%\cite{Yan:2019gbw}
\bibitem{Yan:2019gbw}
S.~F.~Yan, P.~Zhang, J.~W.~Chen, X.~Z.~Zhang, Y.~F.~Cai and E.~N.~Saridakis,
%``Interpreting cosmological tensions from the effective field theory of 
%torsional gravity,''
Phys. Rev. D \textbf{101}, no.12, 121301 (2020)
[\href{\arxiv/arXiv:1909.06388}{arXiv:1909.06388}].
%33 citations counted in INSPIRE as of 22 Sep 2021

%\cite{DAgostino:2020dhv}
\bibitem{DAgostino:2020dhv}
R.~D'Agostino and R.~C.~Nunes,
%Measurements of $H_0$ in modified gravity theories: The role of lensed 
%quasars in the late-time Universe}},
Phys. Rev. D \textbf{101}, no.10, 103505 (2020)
[\href{\arxiv/arXiv:2002.06381}{arXiv:2002.06381}].
%22 citations counted in INSPIRE as of 22 Sep 2021





%\cite{Anagnostopoulos:2020lec}
\bibitem{Anagnostopoulos:2020lec}
F.~K.~Anagnostopoulos, S.~Basilakos and E.~N.~Saridakis,
%Observational constraints on Myrzakulov gravity}},
[\href{\arxiv/arXiv:2012.06524}{arXiv:2012.06524}].



%\cite{Capozziello:2020nyq}
\bibitem{Capozziello:2020nyq}
S.~Capozziello, M.~Benetti and A.~D.~A.~M.~Spallicci,
%Addressing the cosmological $H_0$ tension by the Heisenberg uncertainty}},
Found. Phys. \textbf{50}, no.9, 893-899 (2020)
[\href{\arxiv/arXiv:2007.00462}{arXiv:2007.00462}].
%9 citations counted in INSPIRE as of 22 Sep 2021

%\cite{Saridakis:2019qwt}
\bibitem{Saridakis:2019qwt}
E.~N.~Saridakis, S.~Myrzakul, K.~Myrzakulov and K.~Yerzhanov,
%Cosmological applications of $F(R,T)$ gravity with dynamical curvature and 
%torsion}},
Phys. Rev. D \textbf{102}, no.2, 023525 (2020)
[\href{\arxiv/arXiv:1912.03882}{arXiv:1912.03882}].
%11 citations counted in INSPIRE as of 22 Sep 2021

%\cite{daSilva:2020bdc}
\bibitem{daSilva:2020bdc}
W.~J.~C.~da Silva and R.~Silva,
%``Cosmological Perturbations in the Tsallis Holographic Dark Energy 
%Scenarios,''
Eur. Phys. J. Plus \textbf{136}, no.5, 543 (2021)
%doi:10.1140/epjp/s13360-021-01522-9
[\href{\arxiv/arXiv:2011.09520}{arXiv:2011.09520}].
%5 citations counted in INSPIRE as of 22 Sep 2021




%\cite{Horndeski:1974wa}
\bibitem{Horndeski:1974wa}
G.~W.~Horndeski,
%Second-order scalar-tensor field equations in a four-dimensional space}},
\href{\doi/doi:10.1007/BF01807638}{Int. J. Theor. Phys. \textbf{10}, 363-384} 
(1974)

 


 
 %\cite{DeFelice:2010nf}
\bibitem{DeFelice:2010nf}
A.~De Felice and S.~Tsujikawa,
%Generalized Galileon cosmology}},
Phys. Rev. D \textbf{84}, 124029 (2011)
[\href{\arxiv/arXiv:1008.4236}{arXiv:1008.4236}].



%\cite{Deffayet:2011gz}
\bibitem{Deffayet:2011gz}
C.~Deffayet, X.~Gao, D.~A.~Steer and G.~Zahariade,
%From k-essence to generalised Galileons}},
Phys. Rev. D \textbf{84}, 064039 (2011)
[\href{\arxiv/arXiv:1103.3260} {arXiv:1103.3260}].


%\cite{Renk:2017rzu}
\bibitem{Renk:2017rzu}
J.~Renk, M.~Zumalac\'arregui, F.~Montanari and A.~Barreira,
%Galileon gravity in light of ISW, CMB, BAO and H$_0$ data}},
JCAP \textbf{10}, 020 (2017)
[\href{\arxiv/arXiv:1707.02263}{arXiv:1707.02263}].
%135 citations counted in INSPIRE as of 22 Sep 2021




\bibitem{Ost} M.~Ostrogradsky, %Mémoires sur les équations différentielles, 
%relatives au problème des isopérimètres,
Mem. Acad. St. Petersbourg
{\bf{VI 4}},  385 (1850).



%29 citations counted in INSPIRE as of 08 Oct 20
%\cite{Peirone:2019aua}
\bibitem{Peirone:2019aua}
S.~Peirone, G.~Benevento, N.~Frusciante and S.~Tsujikawa,
%``Cosmological data favor Galileon ghost condensate over $\Lambda$CDM,''
Phys. Rev. D \textbf{100}, no.6, 063540 (2019)
%doi:10.1103/PhysRevD.100.063540
[\href{\arxiv/arXiv:1905.05166}{arXiv:1905.05166}].

 
 
%\cite{Frusciante:2019puu}
\bibitem{Frusciante:2019puu}
N.~Frusciante, S.~Peirone, L.~Atayde and A.~De Felice,
%``Phenomenology of the generalized cubic covariant Galileon model and 
%cosmological bounds,''
Phys. Rev. D \textbf{101}, no.6, 064001 (2020)
%doi:10.1103/PhysRevD.101.064001
[\href{\arxiv/arXiv:1912.07586}{arXiv:1912.07586}].

 

  



%\cite{Kobayashi:2011nu}
\bibitem{Kobayashi:2011nu}
T.~Kobayashi, M.~Yamaguchi and J.~Yokoyama,
%``Generalized G-inflation: Inflation with the most general second-order field 
%equations,''
Prog. Theor. Phys. \textbf{126}, 511-529 (2011)
[\href{\arxiv/arXiv:1105.5723}{arXiv:1105.5723}].


 

%\cite{DeFelice:2011bh}
\bibitem{DeFelice:2011bh}
A.~De Felice and S.~Tsujikawa,
%Conditions for the cosmological viability of the most general 
%scalar-tensor 
%theories and their applications to extended Galileon dark energy models}},
JCAP \textbf{02}, 007 (2012)
[\href{\arxiv/arXiv:1110.3878}{arXiv:1110.3878 }].



%\cite{DeFelice:2010pv}
\bibitem{DeFelice:2010pv}
A.~De Felice and S.~Tsujikawa,
%Cosmology of a covariant Galileon field}},
Phys. Rev. Lett. \textbf{105}, 111301 (2010)
[\href{\arxiv/arXiv:1007.2700}{arXiv:1007.2700}].
%261 citations counted in INSPIRE as of 22 Sep 2021

    
%\cite{Appleby:2011aa}
\bibitem{Appleby:2011aa}
S.~Appleby and E.~V.~Linder,
%The Paths of Gravity in Galileon Cosmology}},
JCAP \textbf{03}, 043 (2012)
[\href{\arxiv/arXiv:1112.1981}{arXiv:1112.1981} ].
%85 citations counted in INSPIRE as of 22 Sep 2021      
 
 %\cite{Babichev:2007dw}
 \bibitem{Babichev:2007dw}
 E.~Babichev, V.~Mukhanov and A.~Vikman,
 %k-Essence, superluminal propagation, causality and emergent geometry}},
 JHEP \textbf{02}, 101 (2008)
 [\href{\arxiv/arXiv:0708.0561}{arXiv:0708.0561}].
 %372 citations counted in INSPIRE as of 22 Sep 2021
 
 %\cite{Deffayet:2010qz}
 \bibitem{Deffayet:2010qz}
 C.~Deffayet, O.~Pujolas, I.~Sawicki and A.~Vikman,
 %``Imperfect Dark Energy from Kinetic Gravity Braiding,''
 JCAP \textbf{10}, 026 (2010)
 [\href{\arxiv/arXiv:1008.0048}{arXiv:1008.0048}].
 %559 citations counted in INSPIRE as of 22 Sep 2021
 
    
%\cite{Saridakis:2010mf}
\bibitem{Saridakis:2010mf}
E.~N.~Saridakis and S.~V.~Sushkov,
%Quintessence and phantom cosmology with non-minimal derivative coupling}},
Phys. Rev. D \textbf{81}, 083510 (2010)
[\href{\arxiv/arXiv:1002.3478}{arXiv:1002.3478}].


%\cite{Capozziello:1999uwa}
\bibitem{Capozziello:1999uwa}
S.~Capozziello and G.~Lambiase,
%``Nonminimal derivative coupling and the recovering of cosmological constant,''
Gen. Rel. Grav. \textbf{31}, 1005-1014 (1999)
[\href{\arxiv/arXiv:gr-qc/9901051}{arXiv:gr-qc/9901051}].
%112 citations counted in INSPIRE as of 20 Sep 2021

%\cite{Koutsoumbas:2013boa}
\bibitem{Koutsoumbas:2013boa}
G.~Koutsoumbas, K.~Ntrekis and E.~Papantonopoulos,
%``Gravitational Particle Production in Gravity Theories with Non-minimal 
%Derivative Couplings,''
JCAP \textbf{08}, 027 (2013)
[\href{\arxiv/arXiv:1305.5741}{arXiv:1305.5741}].
%40 citations counted in INSPIRE as of 20 Sep 2021

%\cite{Feng:2013pba}
\bibitem{Feng:2013pba}
K.~Feng, T.~Qiu and Y.~S.~Piao,
%``Curvaton with nonminimal derivative coupling to gravity,''
Phys. Lett. B \textbf{729}, 99-107 (2014)
[\href{\arxiv/arXiv:1307.7864}{arXiv:1307.7864}].
%40 citations counted in INSPIRE as of 20 Sep 2021


%\cite{Koutsoumbas:2017fxp}
\bibitem{Koutsoumbas:2017fxp}
G.~Koutsoumbas, K.~Ntrekis, E.~Papantonopoulos and E.~N.~Saridakis,
%``Unification of Dark Matter - Dark Energy in Generalized Galileon Theories,''
JCAP \textbf{02}, 003 (2018)
[\href{\arxiv/arXiv:1704.08640}{arXiv:1704.08640}].
%26 citations counted in INSPIRE as of 20 Sep 2021


%\cite{MohseniSadjadi:2013iou}
\bibitem{MohseniSadjadi:2013iou}
H.~Mohseni Sadjadi and P.~Goodarzi,
%``Oscillatory inflation in non-minimal derivative coupling model,''
Phys. Lett. B \textbf{732}, 278-284 (2014)
[\href{\arxiv/arXiv:1309.2932}{arXiv:1309.2932}].

%\cite{Dalianis:2016wpu}
\bibitem{Dalianis:2016wpu}
I.~Dalianis, G.~Koutsoumbas, K.~Ntrekis and E.~Papantonopoulos,
%``Reheating predictions in Gravity Theories with Derivative Coupling,''
JCAP \textbf{02}, 027 (2017)
[\href{\arxiv/arXiv:1608.04543}{arXiv:1608.04543}].
%19 citations counted in INSPIRE as of 20



%\cite{Karydas:2021wmx}
\bibitem{Karydas:2021wmx}
S.~Karydas, E.~Papantonopoulos and E.~N.~Saridakis,
%``Successful Higgs inflation from combined nonminimal and derivative 
%couplings,''
Phys. Rev. D \textbf{104}, no.2, 023530 (2021)
[\href{\arxiv/arXiv:2102.08450}{arXiv:2102.08450}].



   
%\cite{Planck:2018vyg}
\bibitem{Planck:2018vyg}
N.~Aghanim \textit{et al.} [Planck],
%Planck 2018 results. VI. Cosmological parameters}},
Astron. Astrophys. \textbf{641}, A6 (2020)
[erratum: Astron. Astrophys. \textbf{652}, C4 (2021)]
[\href{http://xxx.lanl.gov/abs/1807.06209}
{{\tt arXiv:1807.06209}}].

%\cite{Heisenberg:2022lob}
\bibitem{Heisenberg:2022lob}
L.~Heisenberg, H.~Villarrubia-Rojo and J.~Zosso,
%``Simultaneously solving the $H_0$ and $\sigma_8$ tensions with late dark 
%energy,''
[\href{http://xxx.lanl.gov/abs/2201.11623}
{{\tt arXiv:2201.11623}}].


 

%\cite{Abdalla:2022yfr}
\bibitem{Abdalla:2022yfr}
E.~Abdalla, G.~F.~Abell\'an, A.~Aboubrahim, A.~Agnello, O.~Akarsu, Y.~Akrami, 
G.~Alestas, D.~Aloni, L.~Amendola and L.~A.~Anchordoqui, \textit{et al.}
%``Cosmology Intertwined: A Review of the Particle Physics, Astrophysics, and 
%Cosmology Associated with the Cosmological Tensions and Anomalies,''
[\href{http://xxx.lanl.gov/abs/2203.06142}
{{\tt arXiv:2203.06142}}].
 

 %\cite{Bellini:2014fua}
\bibitem{Bellini:2014fua}
E.~Bellini and I.~Sawicki,
%``Maximal freedom at minimum cost: linear large-scale structure in general 
%modifications of gravity,''
JCAP \textbf{07}, 050 (2014)
[\href{http://xxx.lanl.gov/abs/1404.3713}
{{\tt arXiv:1404.3713}}].


  

%\cite{Peirone:2017ywi}
\bibitem{Peirone:2017ywi}
S.~Peirone, K.~Koyama, L.~Pogosian, M.~Raveri and A.~Silvestri,
%``Large-scale structure phenomenology of viable Horndeski theories,''
Phys. Rev. D \textbf{97}, no.4, 043519 (2018)
[\href{http://xxx.lanl.gov/abs/1712.00444}
{{\tt arXiv:1712.00444}}].

 
 %\cite{Steinhardt:1999nw}
\bibitem{Steinhardt:1999nw}
P.~J.~Steinhardt, L.~M.~Wang and I.~Zlatev,
%``Cosmological tracking solutions,''
Phys. Rev. D \textbf{59}, 123504 (1999).
%doi:10.1103/PhysRevD.59.123504
 
 %\cite{Ezquiaga:2017ekz}
\bibitem{Ezquiaga:2017ekz}
J.~M.~Ezquiaga and M.~Zumalac\'arregui,
%``Dark Energy After GW170817: Dead Ends and the Road Ahead,''
Phys. Rev. Lett. \textbf{119}, no.25, 251304 (2017)
%doi:10.1103/PhysRevLett.119.251304
[arXiv:1710.05901 [astro-ph.CO]].

%\cite{Scolnic:2017caz}
\bibitem{Scolnic:2017caz}
D.~M.~Scolnic, D.~O.~Jones, A.~Rest, Y.~C.~Pan, R.~Chornock, R.~J.~Foley, 
M.~E.~Huber, R.~Kessler, G.~Narayan and A.~G.~Riess, \textit{et al.}
%``The Complete Light-curve Sample of Spectroscopically Confirmed SNe Ia from 
%Pan-STARRS1 and Cosmological Constraints from the Combined Pantheon Sample,''
Astrophys. J. \textbf{859}, no.2, 101 (2018)
[\href{\arxiv/arXiv:1710.00845}{arXiv:1710.00845}].
%943 citations counted in INSPIRE as of 23 Sep 20


%\cite{Jimenez:2001gg}
\bibitem{Jimenez:2001gg}
R.~Jimenez and A.~Loeb,
%``Constraining cosmological parameters based on relative galaxy ages,''
Astrophys. J. \textbf{573}, 37-42 (2002)
[\href{http://xxx.lanl.gov/abs/astro-ph/0106145}
{{\tt arXiv:astro-ph/0106145}}].

 

%\cite{Yu:2017iju}
\bibitem{Yu:2017iju}
 H.~Yu, B.~Ratra and F.~Y.~Wang,
%``Hubble Parameter and Baryon Acoustic Oscillation Measurement Constraints on 
%the Hubble Constant, the Deviation from the Spatially Flat 
%\ensuremath{\Lambda}CDM Model, the Deceleration\textendash{}Acceleration 
%Transition Redshift, and Spatial Curvature,''
Astrophys. J. \textbf{856}, no.1, 3 (2018)
[\href{\arxiv/arXiv:1711.03437}{arXiv:1711.03437}].

\end{thebibliography}
\end{document}